\documentclass[%
aip,
 amsmath,amssymb,
]{revtex4-1}

\usepackage{url}
\usepackage{graphicx}
\usepackage{caption}
\usepackage{subcaption}

\usepackage{dcolumn}
\usepackage{bm}

\usepackage[utf8]{inputenc}
\usepackage[T1]{fontenc}
\usepackage{mathptmx}
\usepackage{amsmath}
\usepackage{wrapfig}
\usepackage{physics}
\usepackage{listings}
\usepackage{xcolor}

\begin{document}

\title{Computational fluid dynamics approach \\
for understanding oscillating and interacting convective flows}

\author{Attila Gergely and Zolt\'an N\'eda}

\email{zoltan.neda@ubbcluj.ro}
\affiliation{Physics Department, Babe\c{s}-Bolyai University, \\
Kogalniceanu street 1, Cluj-Napoca, Romania}

\date{\today}
             
\begin{abstract} 
A 2D numerical hydrodynamics approach is considered for modelling recent experimental results on the oscillation and collective behavior of convective flows. Our simulations 
consider the rising dynamics of heated fluid columns in a gravitational field.  Simulations  are done on two entirely different length-scales, showing also the generality of the investigated phenomena. For the flow of a single heated fluid column, the effect of the inflow yield and the nozzle diameter is studied. In  agreement with the experiments, for a constant nozzle diameter the oscillation frequency increases approximately linearly as a function of the the flow yield and for a constant flow yield the frequency decreases as a power law with the increasing nozzle diameter. Concerning the collective behavior of two nearby flow we find a 
counter-phase synchronization of the oscillations and an increasing trend in the common oscillation frequency when the distance between the flows is decreased. These results are again in agreement with the experimental findings. 
\end{abstract}



\maketitle


\section{Introduction}

Recent experimental results have proven that rising gas columns can perform oscillations and their interaction  leads to fascinating collective behavior \cite{Gergely2021}.  These oscillations have been reported much earlier by Yuan et. al \cite{Yuan1994} and can be discussed in analogy with the very similar phenomena  known for diffusive flames \cite{Chamberlin1948,Durox1995,DUROX1997,Huang1999,Kitahata2009,Ghosh2010,Okamoto2016,Chen2019,Gergely2020}. The toy-model presented in \cite{Gergely2021} is clearly inadequate to explain the fine details observed in our recent experiments, therefore a more sophisticated theoretical approach is needed. The present study contributes in this sense, considering a 2D numerical hydrodynamics approach to this phenomenon.  

For validating the numerical results, the present work considers as reference the experiments realized using a controlled flow of Helium into air \cite{Gergely2021} . In these experiments the Schlieren technique \cite{settles2001schlieren,Leptuch2006} was used to visualize the flow, allowing also a digital processing of the oscillations. From the   images processed by the Otsu method \cite{Otsu-original,Otsu:2020} the characteristic frequency and the relevant synchronization order parameter was derived.  For a better understanding of the phenomena, some sample movies with original recordings and the ones processed with the Otsu method are given in our YouTube channel \cite{Attila3}. For  a single flow column the experiments investigated the effect of the nozzle diameter and flow rate on the observed oscillation frequency.  For the collective behavior of two nearby flows the experiments described in \cite{Gergely2021} investigated the phase difference 
between the oscillations, their frequency and a proper synchronization order parameter as a function of the separation distance between the flows. It was concluded  that at a constant Helium flow yield the oscillation frequency of the rising gas column decreases in from of a  power law as a function of the nozzle diameter. This finding is similar with the observed oscillation frequency of the flames of candle bundles as a function of the number of candles in the bundle \cite{Gergely2020}. For a constant nozzle diameter it was found that the oscillation frequency of the flow increases linearly with the flow yield.  For the collective behavior of two nearby and clearly separated flow columns with similar flow parameters only counter-phase synchronization was observed. This is somehow different from the phenomena observed for candle bundle flames, where both in-phase and counter phase synchronization is present depending on the distance between the flames \cite{Gergely2020}.  The experiments concluded that for short distances the oscillation frequency of the flow column becomes significantly higher than the frequency observed for non-interacting Helium columns with the same parameters (flow rate and nozzle diameter).  All the above summarized results, should be a test for any numerical simulation approach on this interesting phenomenon. 

Due to the complexity of the problems related to flows in different spatial configurations, computation approaches are many times the only theoretical possibilities to realistically model such phenomena (see for example \cite{cfd1,cfd2}). Even with such a modelling methodology, imposing the right boundary conditions and offering a proper discretization of space and time raises many technical challenges \cite{Icase}. The incredible revolution we experience nowadays in computational resources and methods, helped us to overcome much of these difficulties and computational fluid dynamics became the primary tool to investigate theoretically problems related to fluid dynamics. However, even with the presently available computational power, many times we are forced to investigate a simpler flow topology and reduce the dimensionality of the problem \cite{Correa2019}. This is nowadays a standard procedure for those cases where the problem becomes computationally difficult in 3D.   Usually a two-dimensional  simplification is considered when the periodicity and symmetry of the considered flow allows this. Assuming in the followings a cylindrical symmetry for the flow, we consider a two-dimensional numerical fluid dynamics approach to the above described phenomenon.  First, we discuss the theoretical background on which our approach is build and the details of the applied numerical method. Using  simple and straightforward examples we test the simulation environment and gain confidence in the method. After this methodological part, we approach the proposed problem and compare critically the results of the simulations with the experimental data from \cite{Gergely2021}. Finally, conclusions are drown and universality features of this 
intriguing phenomenon is discussed.   

\section{The numerical approach}

We present here a 2D numerical approach, that is suitable 
for modelling the oscillations and collective behavior observed in rising gas columns.  
In order to further simplify the problem, instead of a Helium column injected from the bottom we consider the flow of the same incompressible fluid as the surrounding, heated in a restricted region at the bottom of the simulated area. In such manner we get a rising gas column which is also realizable in experiments.

\begin{figure}
	\centering 
	\includegraphics[width=.75\textwidth]{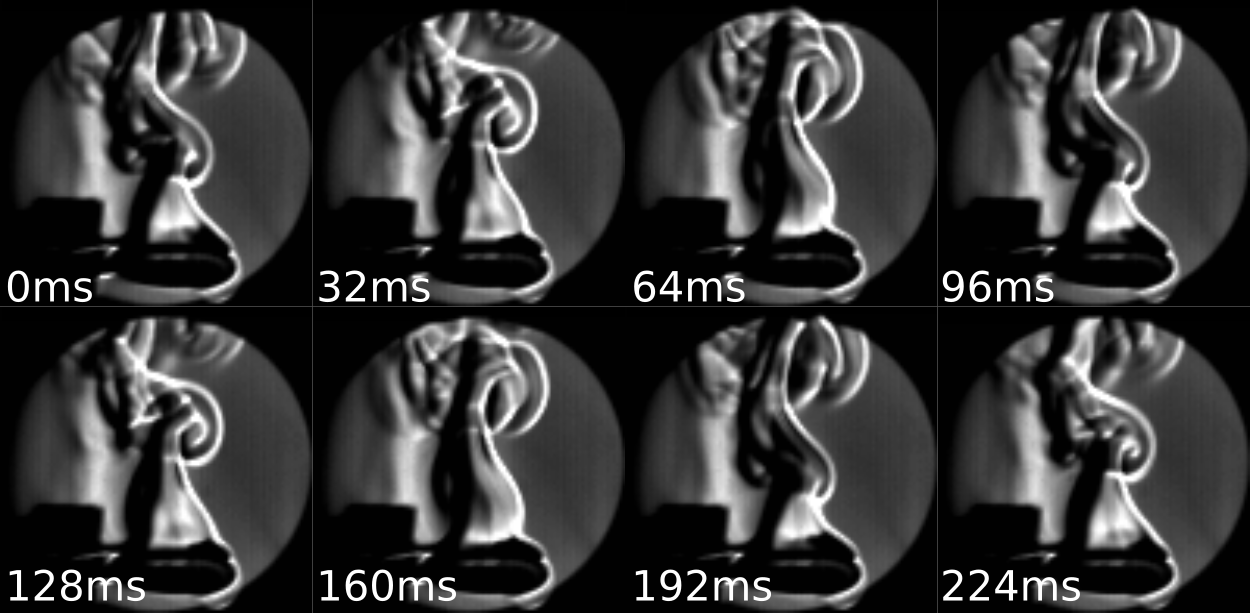}
	\caption{Visualization of a rising hot-air column by the Schlieren technique. Similar instabilities and oscillations appear as in the case of the rising Helium gas column.}
	\label{fig:Schliren-coil-2}
\end{figure}

Using the same Schlieren technique as before, on Figure \ref{fig:Schliren-coil-2} and on the movies presented in \cite{Attila4} we show that very similar instabilities and oscillations occur. In these kitchen-experiments the heating is realized by a simple heating coil in which one controls the dissipated electric power.  Unfortunately in such experiments there is no good control over the flow debit, therefore one cannot conduct such carefully monitored experiments as were done for Helium. This is the main reason why the numerical results are compared with our earlier experiments presented in \cite{Gergely2021}.

The advantage of the proposed setup is that we do not have to apply the numerical fluid dynamics method for two component gases.
We pay however for this simplification by the non-homogeneous temperature field,  therefore extra transports and gradients has to be taken into account. 

In our approach the  fluid is considered to be an ideal one described by the Navier-Stokes equation.  For an incompressible fluid in a gravitational field, the Navier-Stokes equation writes in the following form:
\begin{equation}
	\begin{split}
		 \rho\cdot\frac{\partial \vb{u}}{\partial t}+\rho\cdot\left(\vb{u}\cdot\nabla\right)\cdot\vb{u}&=-\nabla p+\vb{g}\cdot \rho+\mu\cdot \Delta \vb{u}\\
		 \nabla \vb{u}&=0
	\end{split}
	\label{NS}
\end{equation}
Here $\rho$ denotes the density, $p$ is the pressure, $\vb{g}$ the gravitational acceleration, $\vb{u}$ is the velocity of the fluid, and $\mu$ denotes the fluid's viscosity. The quantities $\vb{u}$ and $p$ can be time and position dependent in the flow-space.

For the considered problem, the convective flow due to the temperature difference plays a key role, therefore in the Navier-Stokes equation, we have to take into account the temperature dependence of the density and describe also the time evolution of the temperature inside the fluid. The evolution of temperature and the temperature dependence of the density, are  considered by the following equations:
\begin{equation}
	\begin{split}
		\rho=\frac{\rho_0}{1+(T-T_0)\cdot \alpha} \\
		\frac{\partial T}{\partial t}=D\cdot\Delta T-(\vb{u}\cdot \nabla)\cdot T
	\end{split}
	\label{temp_eq}
\end{equation}
In the above equations, $\rho_0$ is the density at $T_0$, $T$ is the temperature of the fluid at a given spatial position and in a given time-moment,  $D$ is the diffusion constant and $T_0$ is the ambient temperature.
The numerical solution of the coupled systems of partial differential equations (\ref{NS})-(\ref{temp_eq}) was done by using  the "FEniCS" software package \cite{LangtangenLogg2017}. FEniCS is an open-source platform developed for solving Partial Differential Equation (PDE) systems. We chose this platform because it has high-level programming interfaces (C ++, Python), the shape of the equations in the program code is similar to their symbolic form and the program is optimized for a wide range of hardware from laptops to high-performance clusters.

\subsection{The simulation code}
FEniCS uses finite element methods to solve PDEs. As an example in Appendix A we illustrate how to solve the simple 2D Poison equation with FEniCS. 
For our specific problem we first deal with the term describing the evolution of temperature:
\begin{equation}
	\begin{split}
		\frac{\partial T}{\partial t}=D\cdot\Delta T-(\vb{u}\cdot \nabla)\cdot T
	\end{split}
\end{equation}
This equation contains a time derivative, so in addition to the coordinates we also have to discretize time.  This is done by the Euler method as follows:
\begin{equation}
	\begin{split}
		\frac{T(t+dt)-T(t)}{d t}=D\cdot\Delta T(t)-(\vb{u}\cdot \nabla)\cdot T(t),
	\end{split}
	\label{discreteT}
\end{equation}
where the indexes for the $T$ temperature label the discretized time-steps. We then bring each term to the left hand side of the equation, we multiply the equation by a $\tau$ test function, and integrate the equation over the entire simulated domain:
\begin{equation}
	\int_{\Omega}[T(t+dt)-T(t)-D\cdot\Delta T(t)\cdot dt+(\vb{u}\cdot \nabla)\cdot T(t)\cdot dt]\cdot  \tau \  d \Omega=0
	\label{diffT}
\end{equation}
The equations from above contains a second-order derivative for the coordinates, which is eliminated by partial integration:
\begin{equation}
	\int_{\Omega} (\nabla^2\cdot T(t))\cdot \tau \ d\Omega=\int_{\partial \Omega}\left(\frac{\partial T(t)}{\partial \vb{n}}\right)\cdot \tau \ ds-\int_{\Omega} \nabla T(t)\cdot\nabla \tau\ d\Omega
\end{equation}
Here we denoted by $\vb{n}$ the unit normal vector to the $\partial \Omega$ surface. The derivative in respect to $\vb{n}$ is defined as:
\begin{equation}
\frac{\partial T}{\partial \vb{n}} = (\nabla T) \cdot \vb{n}
\end{equation}
Rewriting equation (\ref{diffT}) using the above result and the fact that under the Dirichlet and free boundary conditions the surface integral disappears, we obtain the final form:
\begin{equation}
	\int_{\Omega} ( [T(t+dt)-T(t)+(\vb{u}\cdot \nabla)\cdot T(t) \cdot dt]\cdot \tau\  + D \cdot \nabla T(t) \cdot\nabla \tau \cdot dt\ ) \, d\Omega=0
\end{equation}
The incompressible Navier-Stokes equation (\ref{NS}) was solved using the IPCS (Incremental Pressure Correction Scheme) scheme \cite{LoggMardalEtAl2012}.
The IPCS method consists of 3 steps but before specifying the steps we introduce the following functions and notation:
\begin{equation}
	\begin{split}
		[ \varepsilon (\vb{u}) ]=\frac{1}{2}\cdot ([\nabla \otimes \vb{u}]+[\nabla \otimes \vb{u}]^{T})\\
		[ \sigma (\vb{u},p)] =2\cdot \mu\cdot \varepsilon(\vb{u})-p\cdot I\\
		{\langle \vb{f}, \vb{g}\rangle}_{\Omega}=\int_{\Omega} \vb{f}\cdot \vb{g}\ d\Omega \\
		 {\langle [A], [B]\rangle}_{\Omega}=\int_{\Omega} [A] : [B] \ d\Omega	
	 \end{split}
\end{equation}
The $\otimes$ product is defining a matrix with the following elements:
\begin{equation}
\nabla \otimes \vb{u}=\left[ \frac{\partial u_j}{\partial x_i}  \right]
\end{equation}
We denoted by $[...]$ a square matrix,  by $[...]^T$ the transpose of a matrix and by $:$ the inner product of matrices:
\begin{equation}
[A]:[B] \equiv \sum_{i,j} A_{ij} \, B_{ij}
\end{equation}
Using the $\varepsilon$ and $\sigma$ functions and the specified notation, the steps of the method will be described in the followings. First we reconsider the Navier-Stokes equation in a discretised form on the $\Omega$ velocity space:
\begin{eqnarray}
		&\rho\cdot {\left\langle \frac{\vb{u}^*-\vb{u}(t)}{dt},\vb{v}\right\rangle }_{\Omega}+\rho\cdot{\left\langle \vb{u}(t)\cdot \nabla \vb{u}(t),\vb{v}\right\rangle }_{\Omega}+{\left\langle \left[ \sigma \left(\frac{\vb{u}(t)+\vb{u}^*}{2},p(t)\right) \right] , \left[ \varepsilon (\vb{v}) \right]\right\rangle }_{\Omega}+\\ \nonumber
		&+{\left\langle p(t) \cdot \vb{n},\vb{v}\right\rangle }_{\partial \Omega}-{\left\langle \vb{n} \cdot [\nabla \otimes \left(\frac{\vb{u}(t)+\vb{u}^*}{2}\right)]^{T},\vb{v}\right\rangle }_{\partial \Omega}=\rho\cdot{\left\langle \vb{g},\vb{v}\right\rangle}_{\Omega}
\end{eqnarray}
Here $\vb{v}$ is a test function, for more information on choosing this one should consult 
\cite{LangtangenLogg2017} The first step of the method is the calculation of an intermediate velocity $\vb{u}^*$ from which the pressure will be determined. Then, the pressure is determined in the i-th step in equation:
\begin{equation}
	\begin{split}
		{\langle \nabla p(t+1),\nabla q \rangle }_{\Omega}={\langle \nabla p(t),\nabla q \rangle }_{\Omega}-\frac{{\langle \nabla \vb{u}^*,q\rangle }_{\Omega}}{dt}\cdot \rho
	\end{split}
\end{equation}
In the last step, the velocity in the $t+dt$ time step is determined based on the pressure and the intermediate velocity:
\begin{equation}
	\begin{split}
		{\langle \vb{u}(t+dt),\vb{v}\rangle }_{\Omega}={\langle \vb{u}^*,\vb{v}\rangle }_{\Omega}-\frac{dt\cdot {\langle \nabla (p(t+dt)-p(t)),\vb{v}\rangle }_{\Omega}}{\rho}
	\end{split}
\end{equation}
In the equation from above $q$ is a test function for the pressure.
The method described above for solving the incompressible Navier-Stokes equation is implemented in  2D. In order to solve the equations numerically, we need boundary conditions in addition to discretization.  We used Dirichlet and free boundary conditions. The Dirichlet boundary condition means that the value of the quantity at a given point is fixed. In the case of the free boundary condition, the derivative as a function of the coordinates of the quantity at the given point is 0.

We tested visually our 2D simulation environment in two simple problems. First we intended to reproduce the  Karman vortices in the flow of a fluid around an obstacle (Appendix B).  Second, we simulated the expansion and rising of a heated sphere, verifying the code for non-homogeneous temperature conditions as well (Appendix C). The test simulations reproduced the expected realistic behavior for 
these known problems, giving confidence for the correct implementation of the relevant equations discussed above. 

\subsection{Simulating the rising hot air column}\label{ch:Results}
In the followings we give the details for implementing the simulations aiming to reproduce the characteristic oscillations observed in a rising gas column. The  boundary conditions introduced for velocity and temperature will be justified, and we explain how the time series of the characteristic oscillations were obtained and how the oscillation frequency was calculated.

\begin{figure}
	\centering 
	\includegraphics[width=0.5\textwidth]{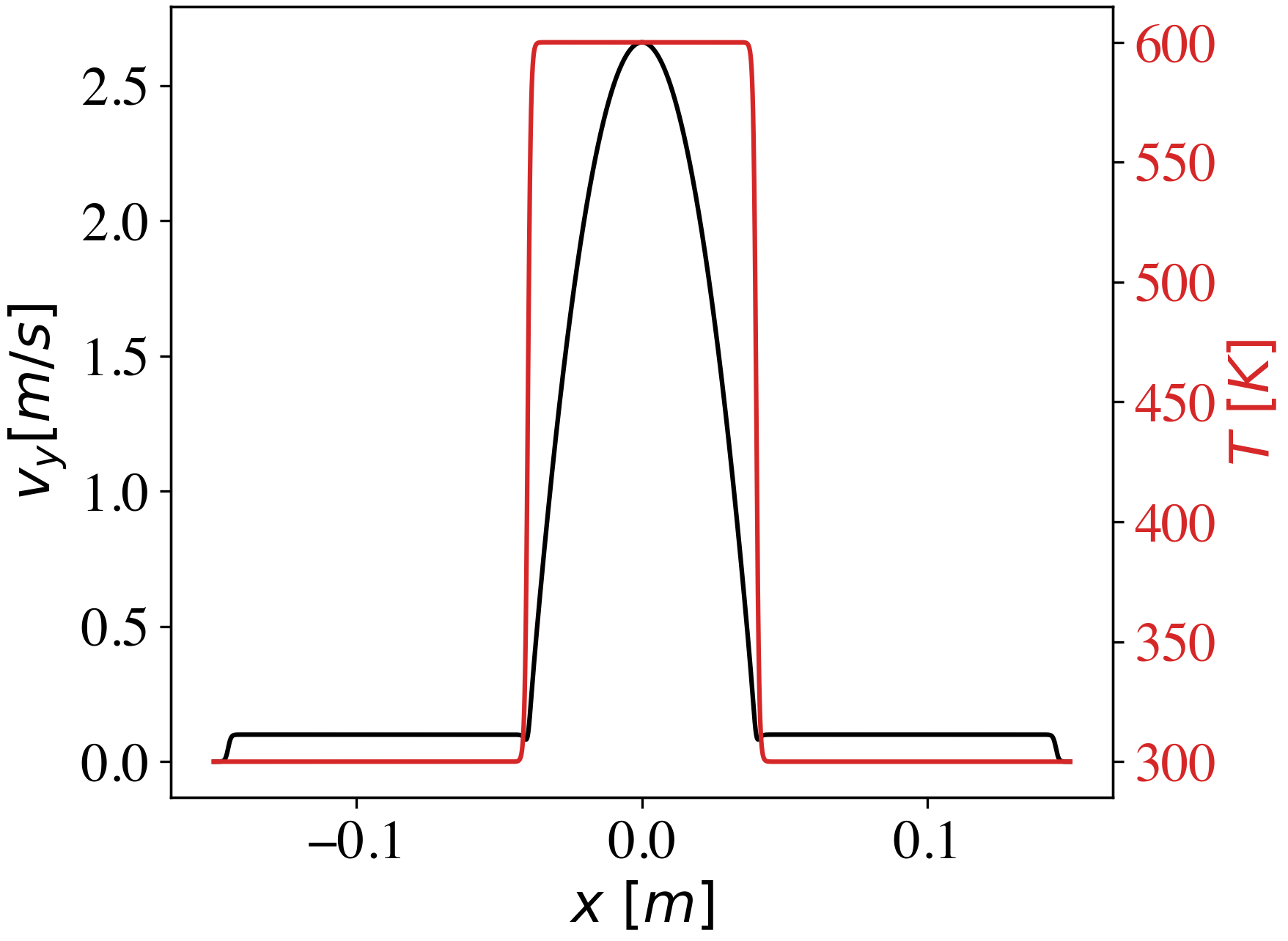}
	\caption{An example for the y-component of velocity, $v_y(x,0)$, and temperature profile, $T(x,0)$, of the heated fluid column at the bottom boundary of the simulated space. The following parameters were used: $c_1=1600\  m^{-1}\cdot s^{-1},\ c_2=0.1\ m\cdot s^{-1},\ c_3=2000\ m^{-1},\ d=0.08\ m,\ T_0=300\, K,\ H=0.3\ m$}
	\label{fig:profil}
\end{figure}

\begin{figure}
	\centering
	\begin{subfigure}[b]{0.52\textwidth}
		\centering
		\includegraphics[width=\textwidth]{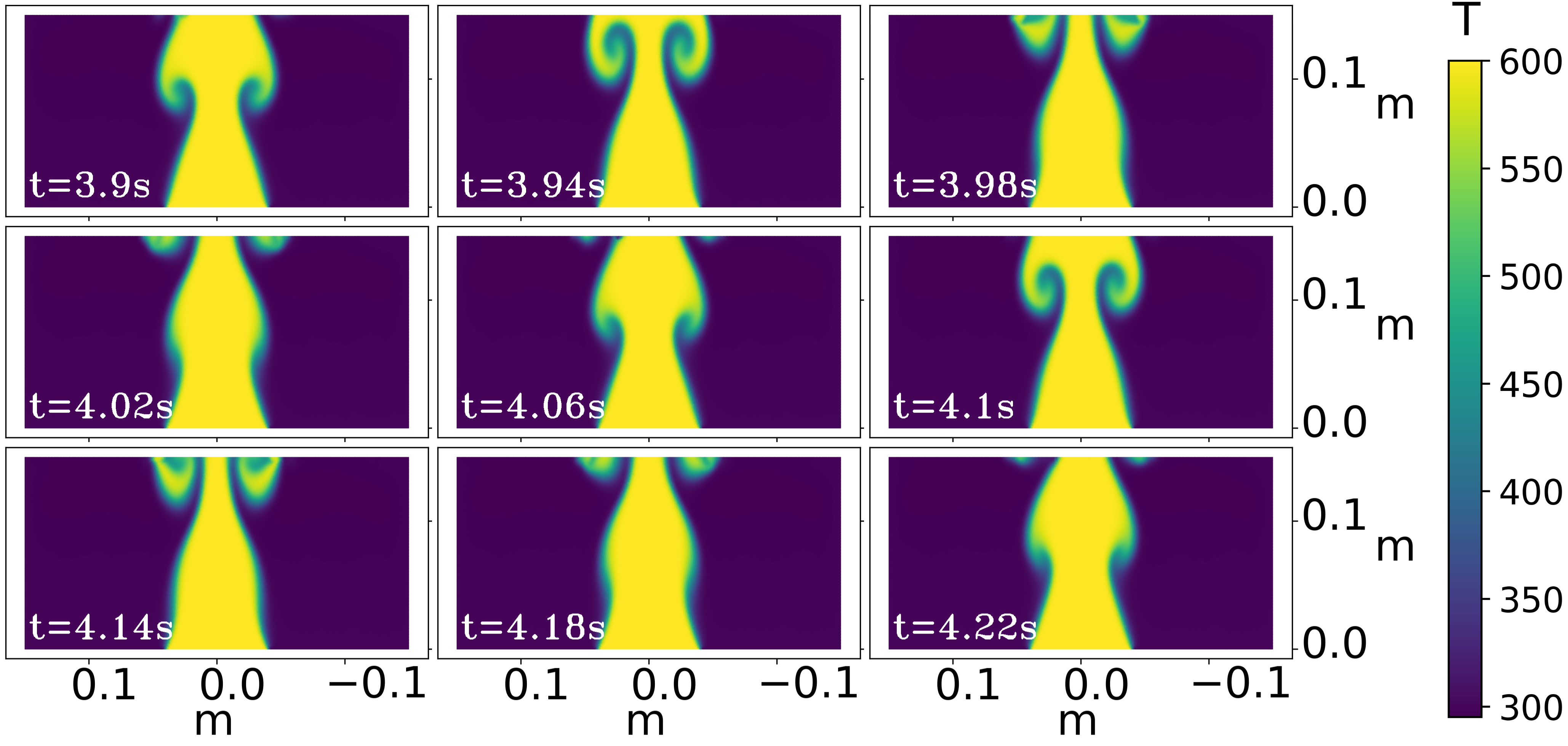}
		\caption{}
		\label{fig:oscDemo_a}
	\end{subfigure}
	\begin{subfigure}[b]{0.4\textwidth}
		\centering
		\includegraphics[width=\textwidth]{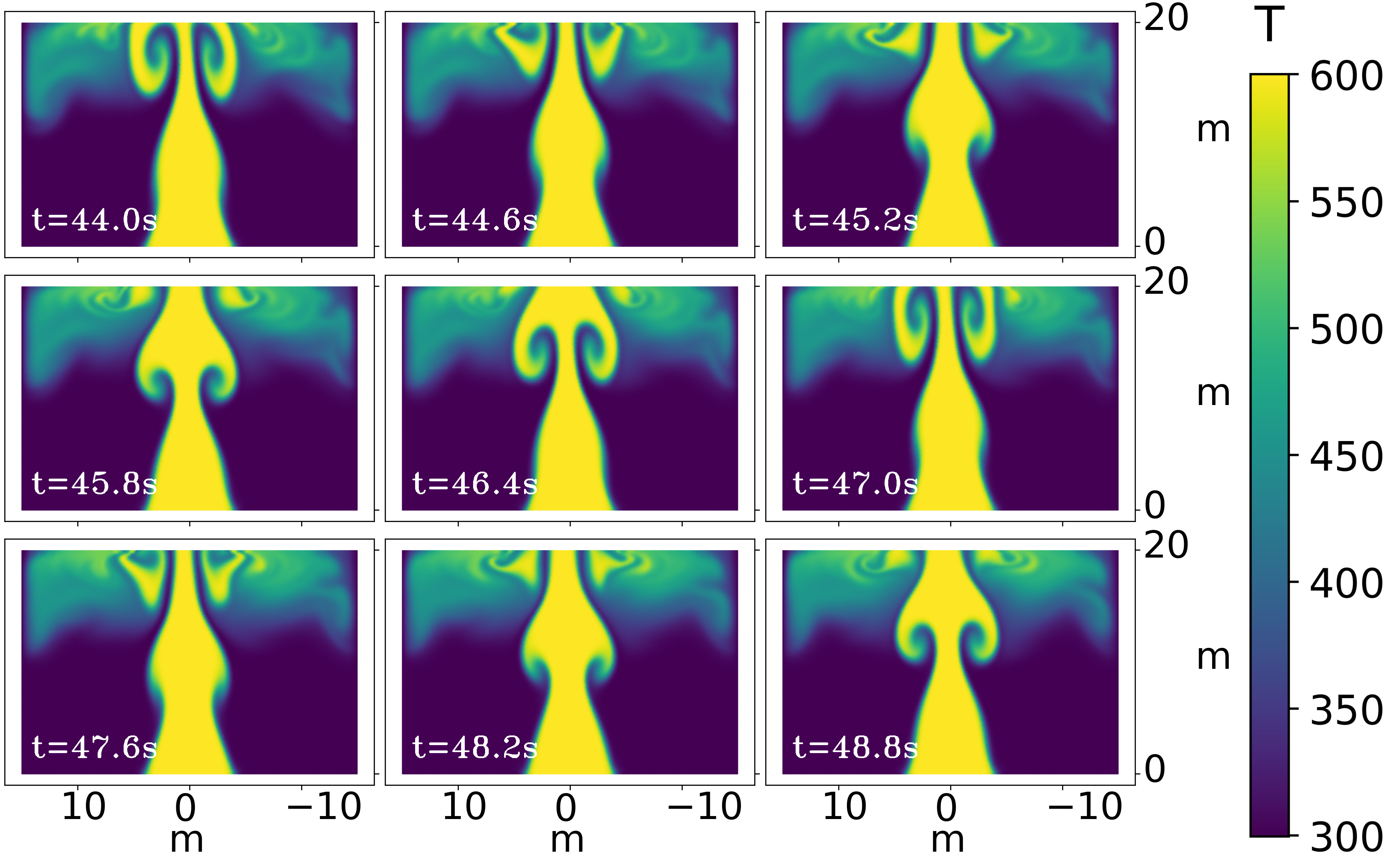}
		\caption{}
		\label{fig:oscDemo_b}
	\end{subfigure}
	\caption{Oscillation of a heated air column in snapshots. The images show the temperature space at the specified $t$ time moments for two different length scales (Figures a. and b.). The gravity acts in the negative direction of the $y$-axis and the parameters of the simulation were chosen as follows:
	(a)$\ \alpha=0.33\cdot 10^{-2}\ K^{-1},\, \rho_0=1.2\ Kg\cdot m^{-2},\, T_0=300\ K,\, D=10^{-4}\ m^2\cdot s^{-1}\cdot K^{-1},\, g_y=-9.81\ m\cdot s^{-2},\, \mu=1.96\cdot10^{-5}\ Kg\cdot s^{-1},\, c_2=0.1\ m\cdot s^{-1},\, c_3=2000\ m^{-1},\, d=0.08\ m,\, c_1=1600\ m^{-1}\cdot s^{-1}$, \\
	(b) $\alpha=10^{-3}\ K^{-1},\, \rho_0=1\ Kg\cdot m^{-2},\, T_0=300\ K,\, D=5\cdot10^{-2}\ m^2\cdot s^{-1}\cdot K^{-1},\, g_y=-9.81\ m\cdot s^{-2},\, \mu=5\cdot10^{-2}\ Kg\cdot s^{-1},\, c_2=0.4\ m\cdot s^{-1},\, c_3=5\ m^{-1},\, d=8\ m,\, c_1=0.375\ m^{-1}\cdot s^{-1}$ }
	\label{fig:oscDemo}
\end{figure}

We consider the inflow geometry presented in Figure \ref{fig:profil} leading to the flow illustrated in Figure \ref{fig:oscDemo} (a). On the sidewalls, the value of the velocity is fixed to 0, on the lower boundary, the $x$-direction component of the velocity is considered as 0 and the $y$-direction component is given by the following parabolic-like kernel (see Figure \ref{fig:profil})
\begin{eqnarray}
		&	v_y(x,0)=c_1f\left(x,\frac{d}{2}\right) \left(\frac{d}{2}-x\right)\left(x+\frac{d}{2}\right)+c_2 f \left(x,\frac{H}{2}\, (1-\frac{1}{30})\right), \nonumber \\
\end{eqnarray}
with:
\begin{eqnarray}
		& f(a,b)=\frac{1}{e^{-c_3\cdot (b+a)}+1}-\frac{1}{e^{c_3\cdot (b-a)}+1}.
\end{eqnarray}

In the equation from above, $d$ denotes the nozzle diameter, $H$ denotes the width of the simulated space, the parameters $c_1, c_2$ determine the incoming flow rate of the fluid, and $c_3$ is a tuning parameter governing the cut in each profile.

At the upper boundary (height $L$), free boundary conditions are applied for the $y$ component of the velocity and for the $x$ component Dirichlet condition is applied, i.e. $v_{x}(x,L)=0$.
For pressure, Dirichlet boundary conditions were used in the upper part of the simulated volume, $p(x,L)=g\cdot \rho_0\cdot l$, and free boundary condition for the other boundaries.
The temperature on the walls is fixed to $T_0$. On the upper boundary we consider $T_0$ if the $y$-direction component of the velocity is negative, otherwise free boundary conditions are used. The temperature at the lower boundary is determined by using the equation:
\begin{eqnarray}
	&	T(x,0)=T_0+T_{heating} f\left(x,\frac{d}{2}\right)
\end{eqnarray}
Here, the $T_{heating}$ temperature governs the extra heat amount introduced in the system. For simplicity reasons we have used in all the presented results $T_{heating}=T_0$.
In the first attempts at the upper part of the simulated box, free boundary condition were considered for the velocity. However in such cases unexpected instabilities occurred and after a certain time the heated fluid column was pushed on one of the sidewalls. We have examined carefully this phenomenon and concluded that a self-amplifying effect is responsible for its development.
Due to the convective flow, the amount of fluid leaving the simulation box is larger than the volume of fluid flowing into the simulation box through the lower boundary. Since the fluid is incompressible, fluid must flow back into the simulation box through the upper boundary. Since there is always an asymmetry in the profile of the fluid inflow this will slightly deflect the outflowing column. In the direction of the deflection, the inflow area decreases, so the asymmetry in the fluid inflow increases. 
An increase in asymmetry over time will result in the fluid flowing along one of the walls. This is the simple explanation of the observed instabilty.

Two methods were used to eliminate these instabilities. The first method is to flow a fluid of ambient temperature $T_0$ at a constant rate on both sides of the heated air column. Since the flow is two-dimensional, the fluid flowing on a given side can only leave on the same side and this will always provide a minimum distance from the wall for the rising jet. The second method is to allow only the $y$-direction component of the velocity at the upper boundary. Combining this two methods will eliminate  the tendency of the jet to approach one of the sidewalls.

For the upper boundary a proper boundary condition has to be applied also for the inflowing fluid  temperature. At the upper boundary an inflow is also necessary in order to respect the incompressibility of the fluid. Since the temperature of the outflowing fluid varies over a wide range we can't apply Dirichlet boundary condition to the whole upper boundary because this would cause unmanageable gradients. Avoiding large gradients due to large temperature differences was solved by applying boundary condition only to those points where the $y$-direction component of the velocity became negative.

The time series for the relevant hydrodynamical parameters were generated and we followed the temperature distribution in the simulated space.
For realistically chosen parameters, it was shown that the model is capable to produce an oscillation similar to the one observed in the case of the Helium column. Interestingly, 
it was found that such oscillations are possible even on largely different length-scales. 
The observed oscillation is shown in Figure \ref{fig:oscDemo} a, b. where we illustrate the temperature space at subsequent time moments. For the simulations presented in Figure \ref{fig:oscDemo}, we used the parameters spcified in the figure caption. 

For a quantitative evaluation of the simulated dynamics, the Otsu method was applied for the 2D temperature field. To obtain the time series, in uniform time intervals the Otsu processed pixels were summed up to a certain height, after this the obtained time series was divided by the average value of the time series. The oscillation frequency was calculated in a similar manner with the experiments, based on the above generated time series. In the first step a Fourier transform was applied to the time series and then the value of the frequency belonging to the largest peak was determined as the relevant oscillation frequency.

With the implemented simulation code we examined how the inflow rate (yield) of the heated fluid column and the nozzle diameter affects its oscillation frequency. We also investigated the  collective behavior for the oscillation of two columns placed nearby each other.

\subsection{Numerical results for the oscillation frequency}

The effect of flow yield and nozzle diameter was examined on two different length-scales. To study the flow yield we used the parameter sets (a), (b) introduced above and the nozzle diameters were $d=0.08\ m$ and $d=8\ m$ respectively.
For constant $c_3$, $c_2$, and $d$ parameters the yield (flow debit) of the heated fluid depends only on $c_1$:
\begin{eqnarray}
&	\Phi=\int_{-\frac{d}{2}}^{\frac{d}{2}}v_y(x,0)\ dx=  \label{eq:flowyield} \\
&	= \int_{-\frac{d}{2}}^{\frac{d}{2}} [c_1f\left(x,\frac{d}{2}\right) \left(\frac{d}{2}-x\right)\left(x+\frac{d}{2}\right)+c_2 f \left(x,\frac{H}{2}(1-\frac{1}{30})\right) ] dx  \nonumber 
\end{eqnarray}

\begin{figure}
	\centering
	\begin{subfigure}[b]{0.388\textwidth}
		\centering
		\includegraphics[width=\textwidth]{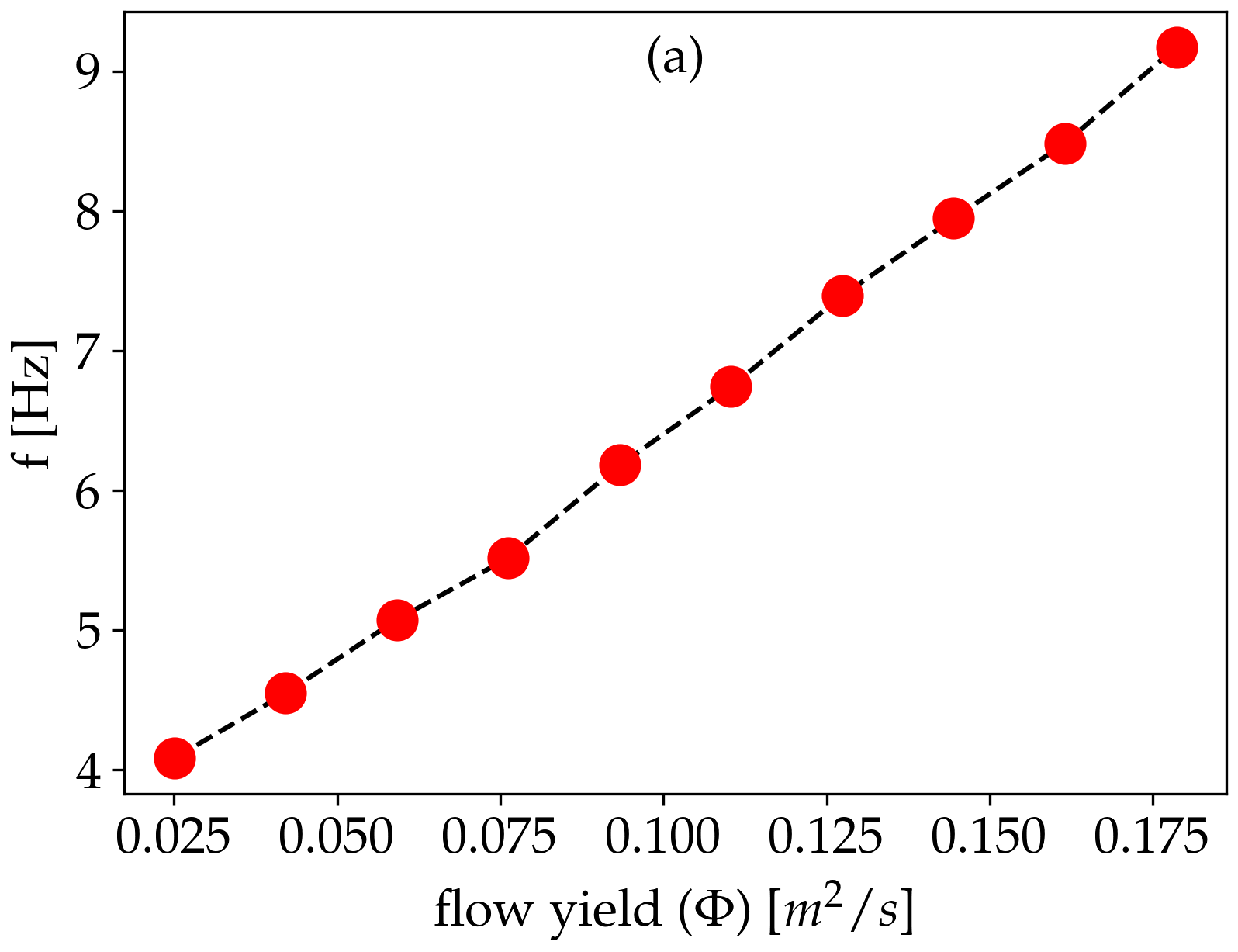}
		\label{fig:flow_y_a}
	\end{subfigure}
	\begin{subfigure}[b]{.4\textwidth}
		\centering
		\includegraphics[width=\textwidth]{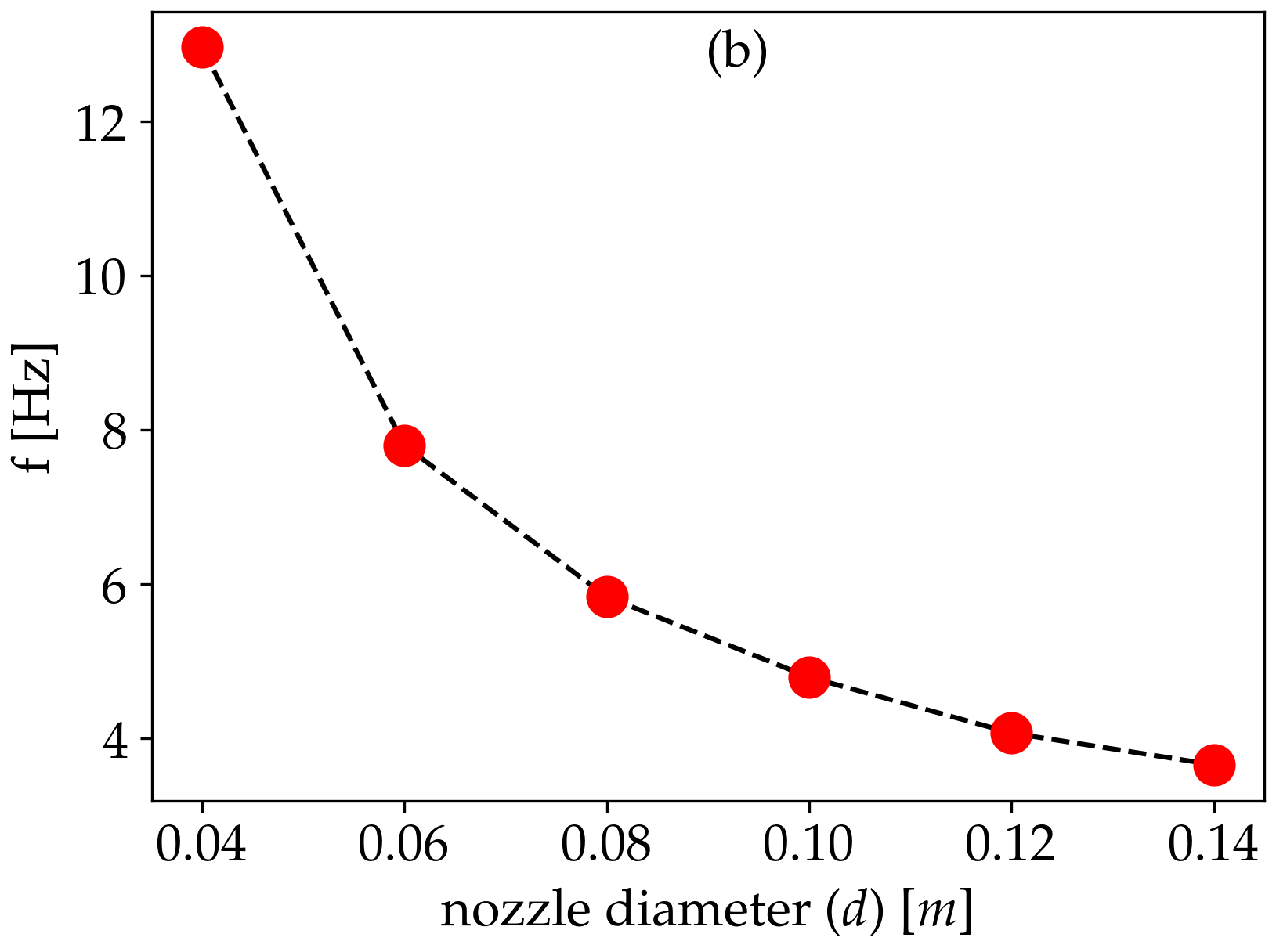}
		\label{fig:diam_b}
	\end{subfigure}
	\caption{Simulation results for the smaller length-scale. Figure (a) shows the oscillation frequency of the heated fluid column as a function of the flow yield $\Phi$ fixed by equation (\ref{eq:flowyield})  for $d\ =\ 0.08\ m$ inflow diameter. Figure (b) shows the oscillation frequency of the heated fluid column as a function of the $d$ nozzle diameter. The other parameters used  are the same as the ones specified in the caption of Figure \ref{fig:oscDemo},  the value of $c_1$ for the different nozzle diameters are given in Table \ref{tab:c_small_scale}.}
	\label{fig:nozzle_diam_flow_real_sys}
\end{figure}

\begin{figure}
	\centering
	\begin{subfigure}[b]{0.45\textwidth}
		\centering
		\includegraphics[width=\textwidth]{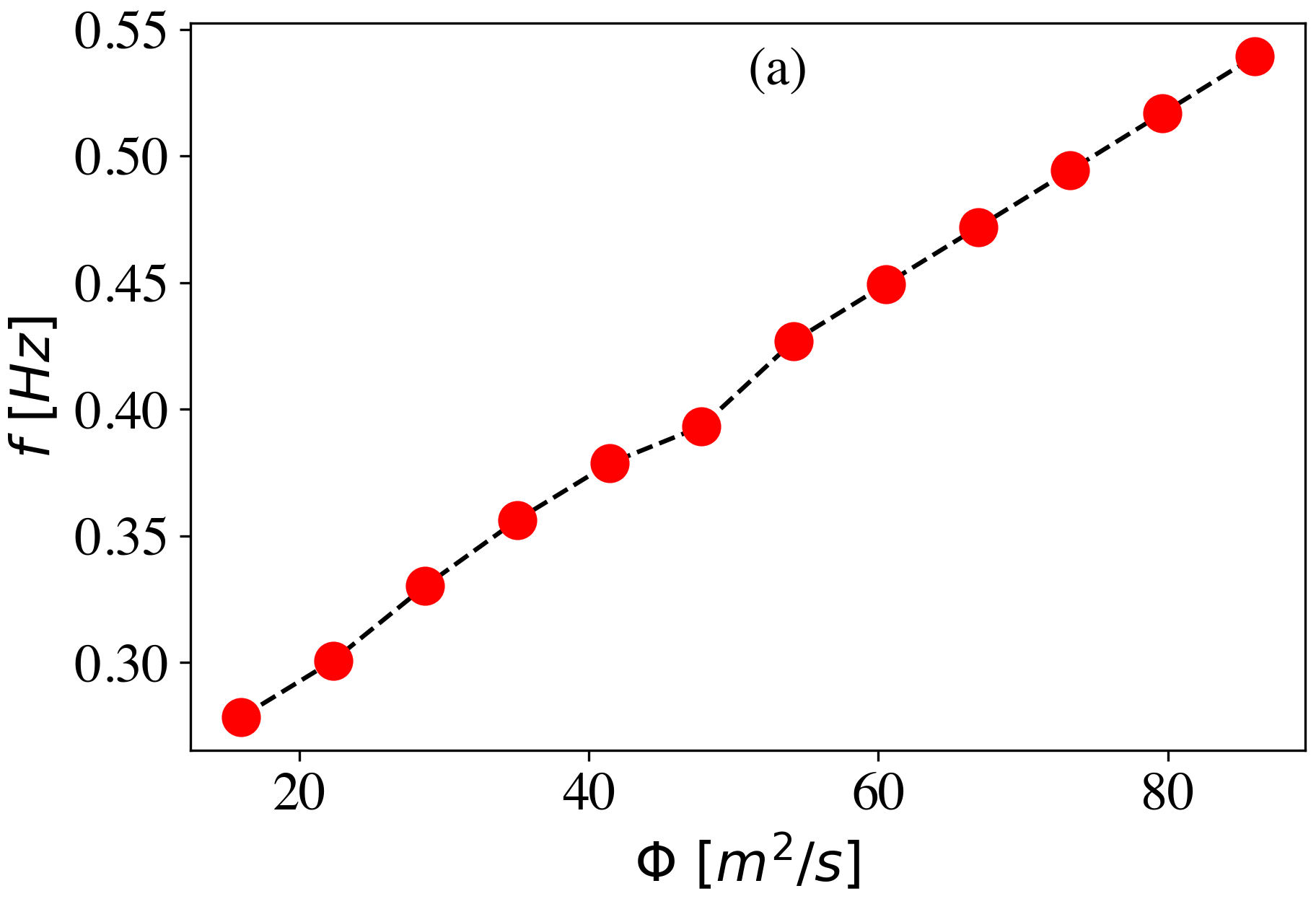}
		\label{fig:y equals x}
	\end{subfigure}
	\begin{subfigure}[b]{0.45\textwidth}
		\centering
		\includegraphics[width=\textwidth]{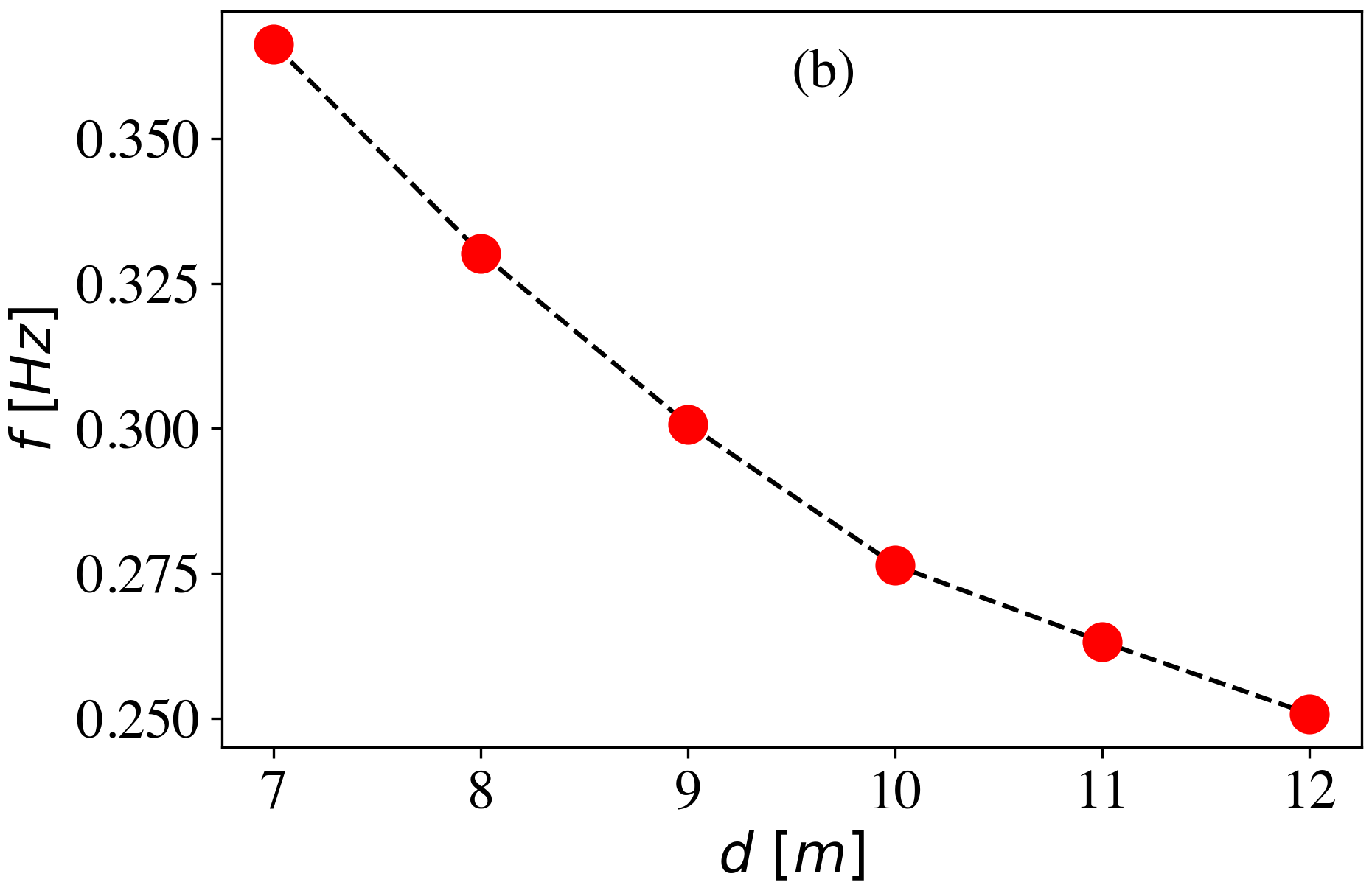}
		\label{fig:three sin x}
	\end{subfigure}
	\caption{Simulation results for the larger length-scale. Figure (a) shows the oscillation frequency of the heated fluid column as a function of the flow yield $\Phi$ fixed by equation (\ref{eq:flowyield})  for $d\ = 8\ m$ inflow diameter. Figure (b) shows the oscillation frequency of the heated fluid column as a function of the $d$ nozzle diameter. The other parameters are the same as the ones specified in the caption of Figure \ref{fig:oscDemo},  the value of $c_1$ for the different nozzle diameters are given in Table \ref{tab:c}.}
	        \label{fig:inflow}
\end{figure}

\begin{figure}
	\centering 
	\includegraphics[width=0.6\textwidth]{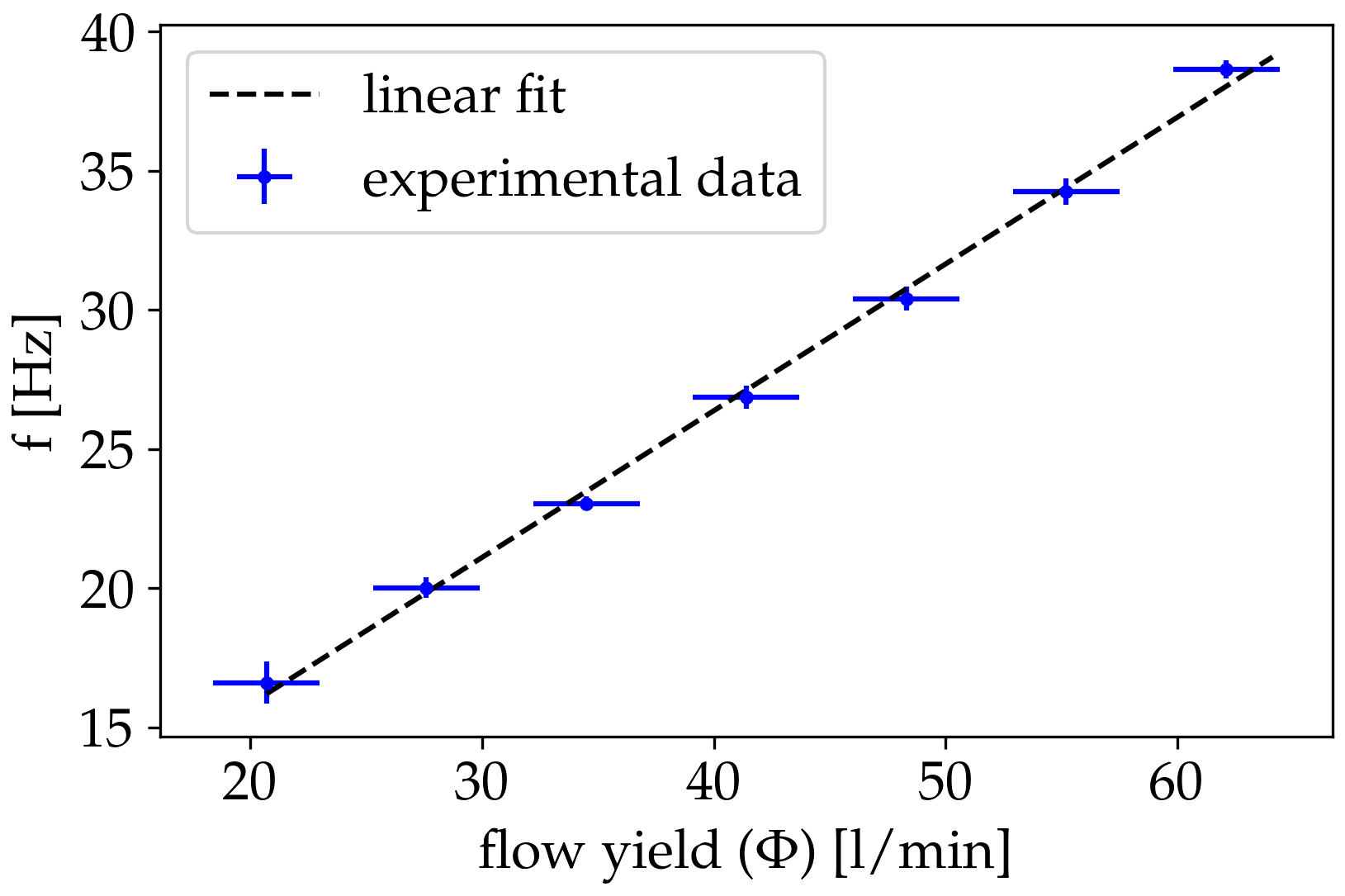}
	\caption{Experimentally observed oscillation frequency of a Helium column as a function of the yield (flow debit) obtained for a setup with nozzle diameter of 2 cm. With increasing flow yield, the frequency of the oscillation increases in an almost linear manner. The plot is done by using the results presented  in \cite{Gergely2021}. Error bars show the uncertainty in the measurement of the flow yield. }
	\label{fig:heliumFreqYield}
\end{figure}

The computed oscillation frequency of the heated fluid column as a function of the $\Phi$ parameter is plotted in Figure \ref{fig:inflow}a and Figure \ref{fig:nozzle_diam_flow_real_sys}a. One will observe that the oscillation frequency increases as the flow rate $\Phi$ increases and this increasing trend can be well approximated by a linear fit in good agreement with the experimental results plotted in Figure \ref{fig:heliumFreqYield}.

The effect of nozzle diameter on the oscillation frequency was investigated at a constant inflow yield.
Since the yield $\Phi$  depends on $d$ according to equation  (\ref{eq:flowyield}), for different nozzle diameters we must rescale the parameters $c_1$ so that the flow rate remains constant.
For the smaller length-scale simulations, we used  $\Phi_1=0.076\ m^2/s$ flow yield and for the larger scale simulations, we used $\Phi_2=29\ m^2/s$ flow yield. To keep constant the flow yield for different nozzle diameters, we varied the value of the $c_1$ parameter. The $c_1$ values for the different nozzle diameters are shown in Table \ref{tab:c_small_scale} and Table \ref{tab:c}.
 \begin{table}[h!]
	\centering
	\begin{tabular}{ |c|c|c|c|c|c|c| }
	\hline
	d\ [m]&0.04 &0.06 &0.08 &0.1 &0.12&0.14 \\
	 \hline
	 $c_1\ [m^{-1}\cdot s^{-1}]$&6781&1952&800&398&223&136\\
	 \hline
	 \end{tabular}
	  \caption{Value of the $c_1$  parameter for different nozzle diameters $d$, in order to keep the flow rate $\Phi_1=0.076\ m^2/s$.}
	  \label{tab:c_small_scale}
 \end{table}
 
\begin{figure}
	\centering 
	\includegraphics[width=0.6\textwidth]{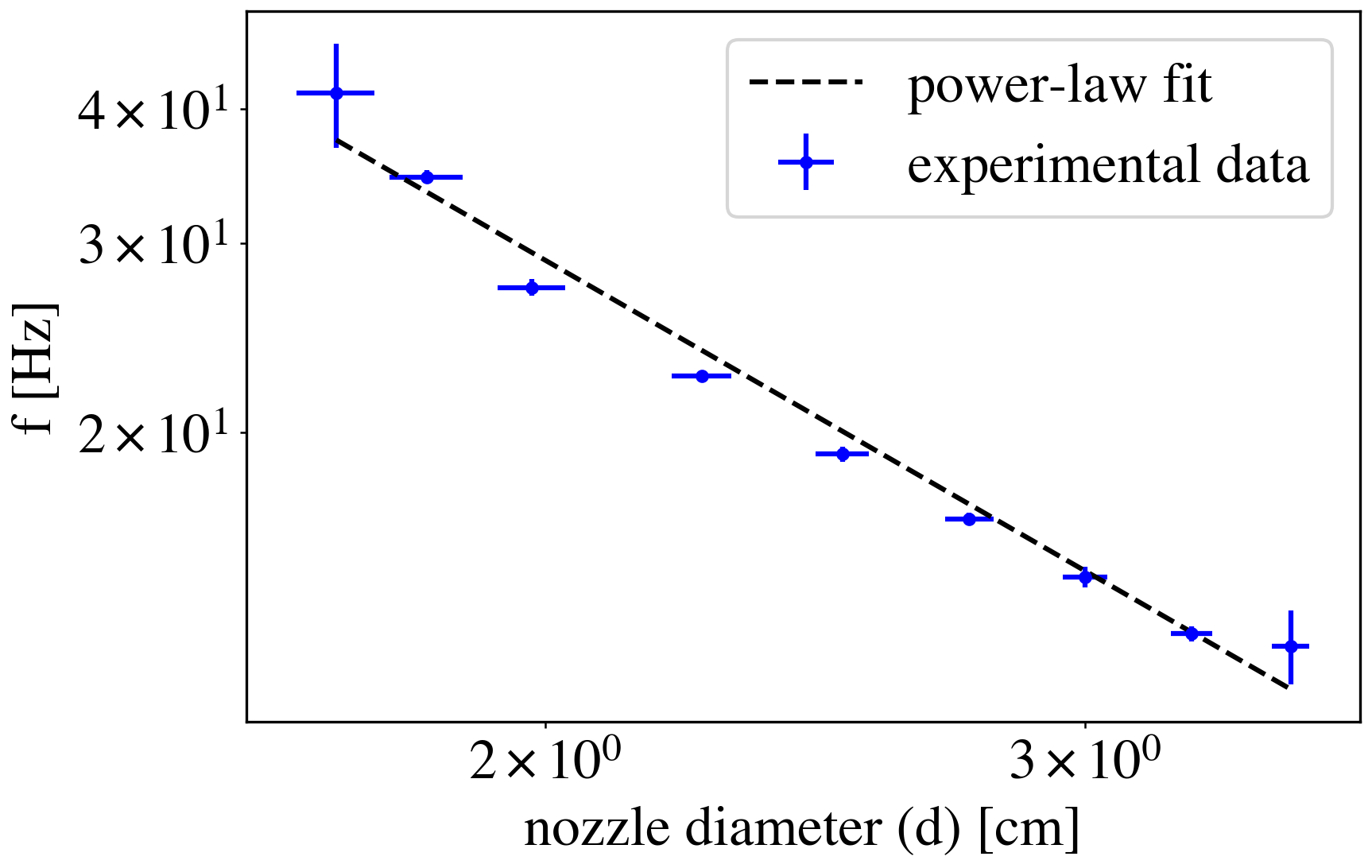}
	\caption{Experimentally observed oscillation frequency of a Helium column as a function of the nozzle diameter 
for a yield of $\Phi=$46 $\pm$ 2.3 $l / min$. The plot is done by using the results presented  in \cite{Gergely2021}. Error bars show the variations of the experimentally used nozzle diameter in different directions. }
	\label{fig:Oscillation frequency nozzle diam experimental}
\end{figure}

\begin{table}[h!]
	\centering
	\begin{tabular}{ |c|c|c|c|c|c|c| }
	\hline
	d\ [m]&7&8&9&10&11&12\\
	 \hline
	 $c_1\ [m^{-1}\cdot s^{-1}]$&0.45&0.3&0.21&0.153&0.11&0.088\\
	 \hline
	 \end{tabular}
	  \caption{Value of the $c_1$  parameter for different nozzle diameters $d$, in order to keep the flow rate $\Phi=29\ m^2/s$.}
	  \label{tab:c}
 \end{table}
For both length scales a decreasing trend of the oscillation frequency  as a function of the nozzle diameter was observed. Results in such sense are plotted in Figure \ref{fig:inflow}b and Figure \ref{fig:nozzle_diam_flow_real_sys}b, the trend 
is in good agreement with the experimental results from Figure \ref{fig:Oscillation frequency nozzle diam experimental}.

\subsection{Numerical results for the collective behavior}

We turn now our interest on reproducing the experimentally observed collective behaviour in form of anti-phase synchronization.

\begin{figure}
	\centering 
	\includegraphics[width=0.75\textwidth]{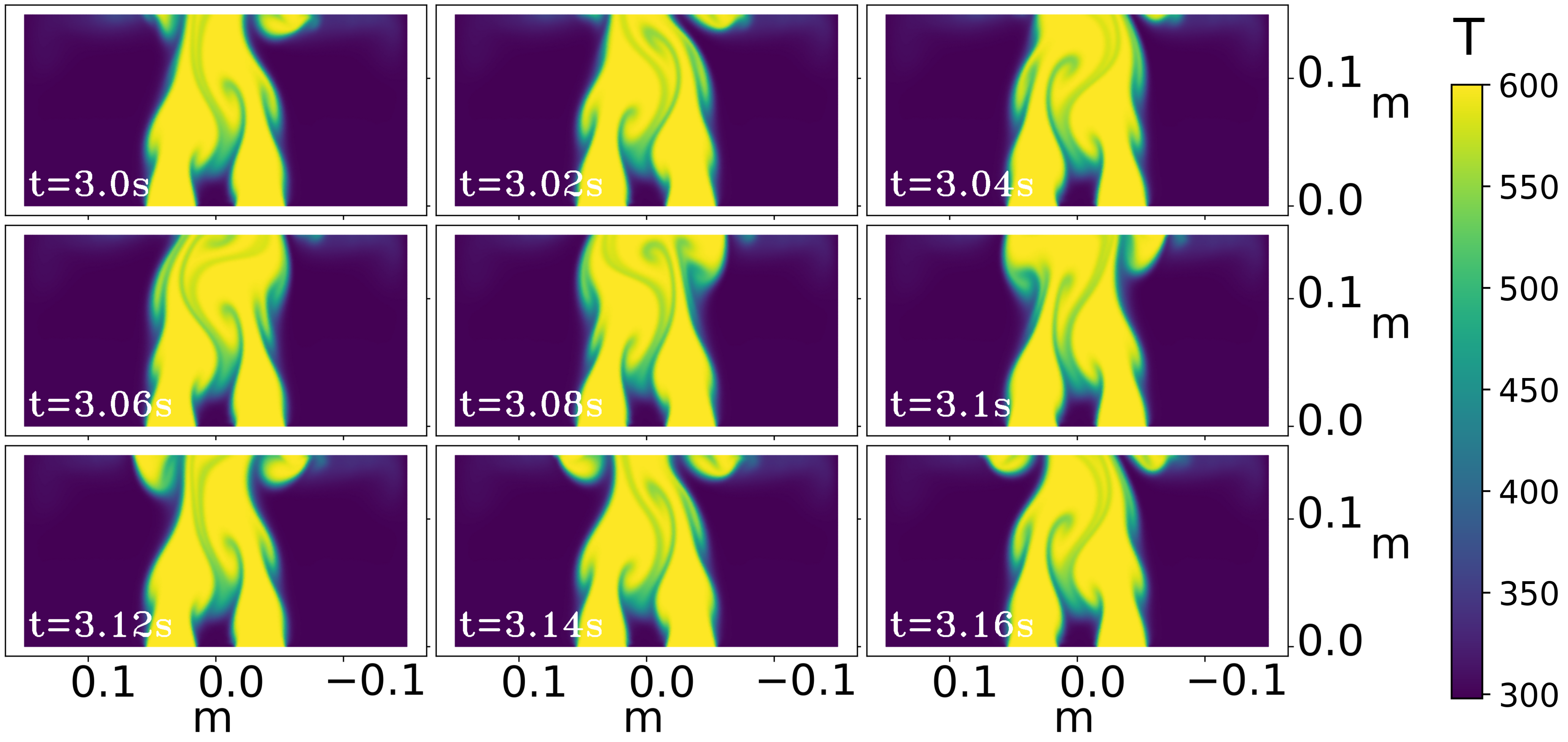}
	\caption{Counter-phase synchronization of two nearby heated columns. Computer simulation results
with the following parameters: 
	$\ \alpha=0.33\cdot 10^{-2}\ K^{-1},\, \rho_0=1.2\ Kg\cdot m^{-2},\, T_0=300\ K,\, D=10^{-4}\ m^2\cdot s^{-1}\cdot K^{-1},\, g_y=-9.81\ m\cdot s^{-2},\, \mu=1.96\cdot10^{-5}\ Kg\cdot s^{-1},\, c_2=0.1\ m\cdot s^{-1},\, c_3=2000\ m^{-1},\, d=0.04\ m,\, c_1=6400\ m^{-1}\cdot s^{-1},\ H=0.3\ m$}
	\label{fig:counter-phase-sync}
\end{figure}

The dimensions of the simulation boxes used to study the collective behavior are the following: 46 m wide (H=46 m) and 30 m high (L=30 m) for the large length-scale and 0.3 m wide (H=0.3 m) and 0.15 m high (L=0.15 m) at the smaller length-scale.
At the lower boundary, the $x$ component of the inflow fluid velocity is 0, and the $y$ component is given by the following kernel function:
\begin{eqnarray}
&		v_y(x,0)=c_1 f\left(x-d_0,\frac{d}{2}\right) \left(d_0-x+\frac{d}{2}\right) \left(x-d_0+\frac{d}{2}\right)+ \nonumber \\
&		+c_1 f\left(x+d_0,\frac{d}{2}\right) \left(\frac{d}{2}-d_0-x\right)\left(x+d_0+\frac{d}{2}\right)
 + c_2 f \left(x,\frac{H}{2}-c_4\right) \nonumber \\
\end{eqnarray}
Here the value of $c_4$ is $0.5\ m$ for the large scale system and $0.005\ m$ for the small scale system.
\begin{figure}
	\centering 
	\includegraphics[width=0.6\textwidth]{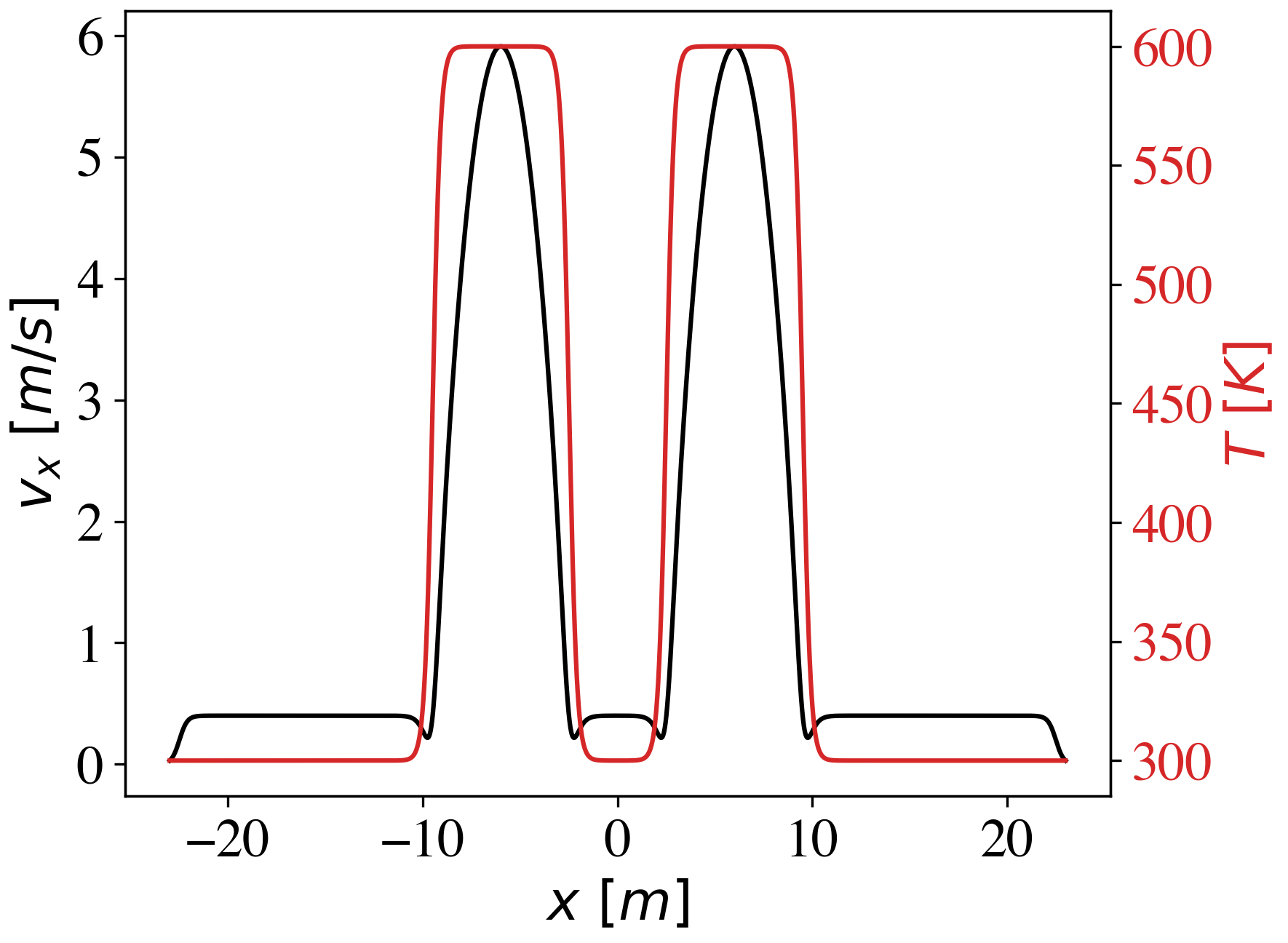}
	\caption{An example for the y-component velocity, $v_y(x,0)$, and temperature, $T(x,0)$, profiles of the heated fluid columns at the bottom boundary of the simulated space. The following parameters were used:$c_1=0.45\  m^{-1}\cdot s^{-1},\ c_2=0.4\ m\cdot s^{-1},\ c_3=5\ m^{-1},\ d=7\ m,\ d_0=6\ m,\ T_0=300\ K,\ H=46\ m$.}
	\label{fig:profilCollective}
\end{figure}
This leads to an inflow profile with two peaks where the centers are separated at a distance of $2d_0$, as it is illustrated in Figure \ref{fig:profilCollective}. The temperature profile is adjusted accordingly:
\begin{eqnarray}
&		T(x,0)=T_0+T_{heating} f\left(x-d_0,\frac{d}{2}\right)+T_{heating} f\left(x+d_0,\frac{d}{2}\right) \nonumber \\  
\end{eqnarray}
We used the same simulation parameters as before and fixed $d=7\ m$, $c_1=0.45\ m^{-1}\cdot s^{-1}$ values for the large length-scale and $d=0.04\ m$, $c_1=6400\ m^{-1}\cdot s^{-1}$ values for the small length-scale system. Again, for the presented results we considered $T_{heating}=T_0$.
Experimental results from \cite{Gergely2021} shows that at small separation distance collective behavior in form of counter-phase synchronization appears. A snapshot for a simulated stable collective behavior is visible in Figure \ref{fig:counter-phase-sync}, reproducing successfully this counter phase synchronization on the smaller length-scale. Similar behavior is observable for the larger length-scales as well. 

\begin{figure}
	\centering
	\begin{subfigure}[b]{0.4\textwidth}
		\centering
		\includegraphics[width=\textwidth]{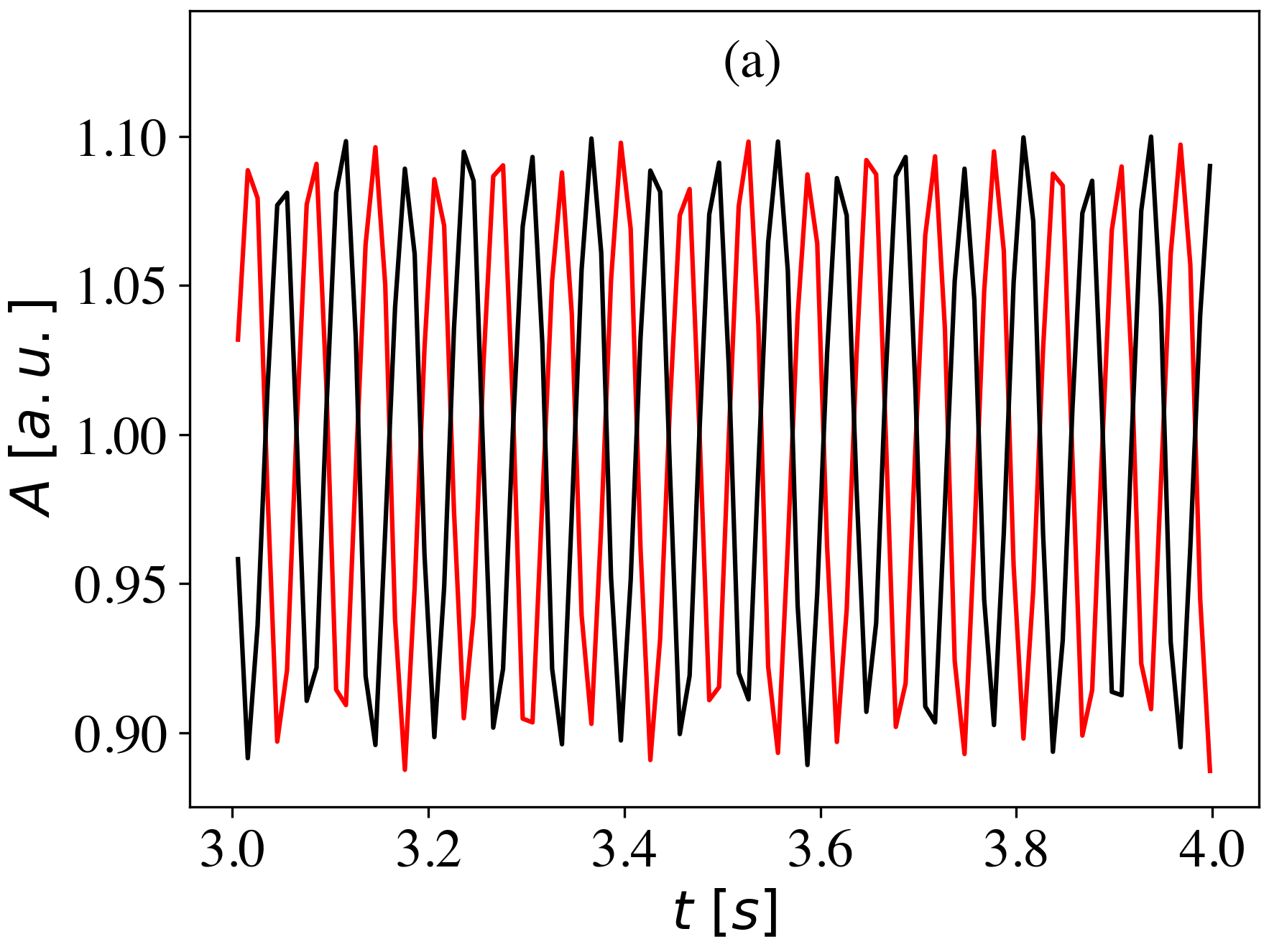}
		\label{fig:colectiveSmal}
	\end{subfigure}
	\begin{subfigure}[b]{0.4\textwidth}
		\centering
		\includegraphics[width=\textwidth]{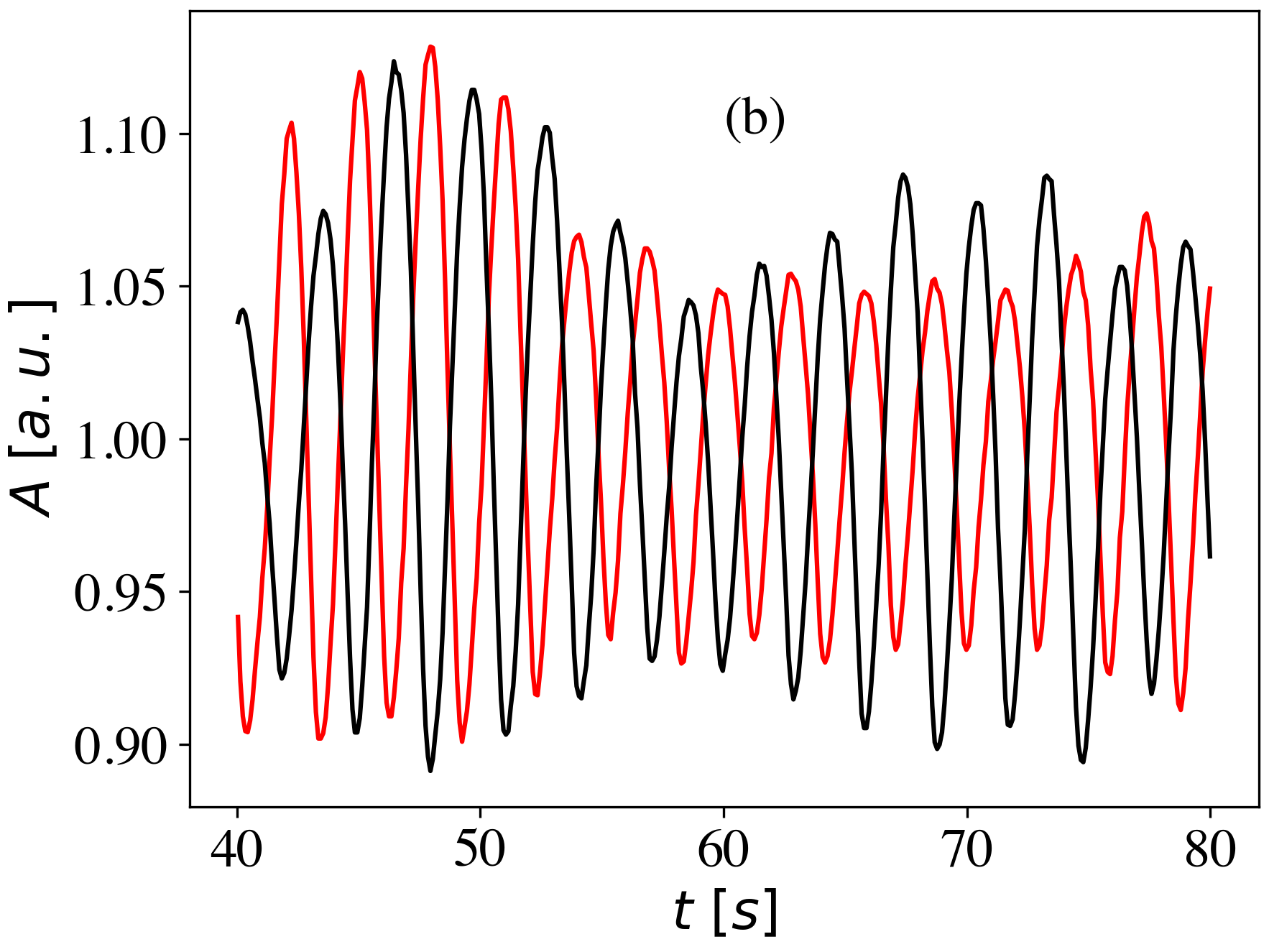}
		\label{fig:colectiveLarge}
	\end{subfigure}
	\caption{Simulated time series for the oscillations of nearby heated fluid columns. 
	Motion of the interface is detected by the Otsu method at the same height from the nozzle.
	For not too high separation distances ((a) $2\cdot d_0=0.03\ m$ , (b)  $2\cdot d_0=5\ m$) a clear counter-phase synchronization is observable. 
	The parameters of the simulations are \\
	(a): $\ \alpha=0.33\cdot 10^{-2}\ K^{-1}$, $\rho_0=1.2\ Kg\cdot m^{-2}$, $T_0=300\ K$, $D=10^{-4}\ m^2\cdot s^{-1}\cdot K^{-1}$,\, $g_y=-9.81\ m\cdot s^{-2}$, $\mu=1.96\cdot10^{-5}\ Kg\cdot s^{-1}$, $c_2=0.1\ m\cdot s^{-1}$, $c_3=2000\ m^{-1}$, $d=0.04\ m$, $c_1=6400\ m^{-1}\cdot s^{-1}$, $2\cdot d_0=0.03\ m$. \\
	(b): $\alpha=10^{-3}\ K^{-1}$,  $\rho_0=1\ Kg\cdot m^{-2}$, $T_0=300\ K$, $D=5\cdot10^{-2}\ m^2\cdot s^{-1}\cdot K^{-1}$, $g_y=-9.81\ m\cdot s^{-2}$, $\mu=5\cdot10^{-2}\ Kg\cdot s^{-1}$,  $c_2=0.4\ m\cdot s^{-1}$, $c_3=5\ m^{-1}$, $d=8\ m$, $c_1=0.375\ m^{-1}\cdot s^{-1}$, $2\cdot d_0=0.03\ m$ and we have fixed $d=7$ and $c_1=0.45\ m^{-1}\cdot s^{-1}$. 	
	}
	\label{fig:HeatedFluis}
\end{figure}

For the pictures processed with the Otsu method the collective oscillation of nearby heated fluid columns  are shown in Figures \ref{fig:HeatedFluis} a and b, for the small and large length-scales, respectively.
For the indicated smaller separation distances an almost perfect counter-phase synchronization develops.  For larger separation distances the phases of the oscillations will begin to shift relative to each other and no clear phase-difference blocking is observable.

\begin{figure}
	\centering
	
	\begin{subfigure}[b]{.4\textwidth}
		\centering
		\includegraphics[width=\textwidth]{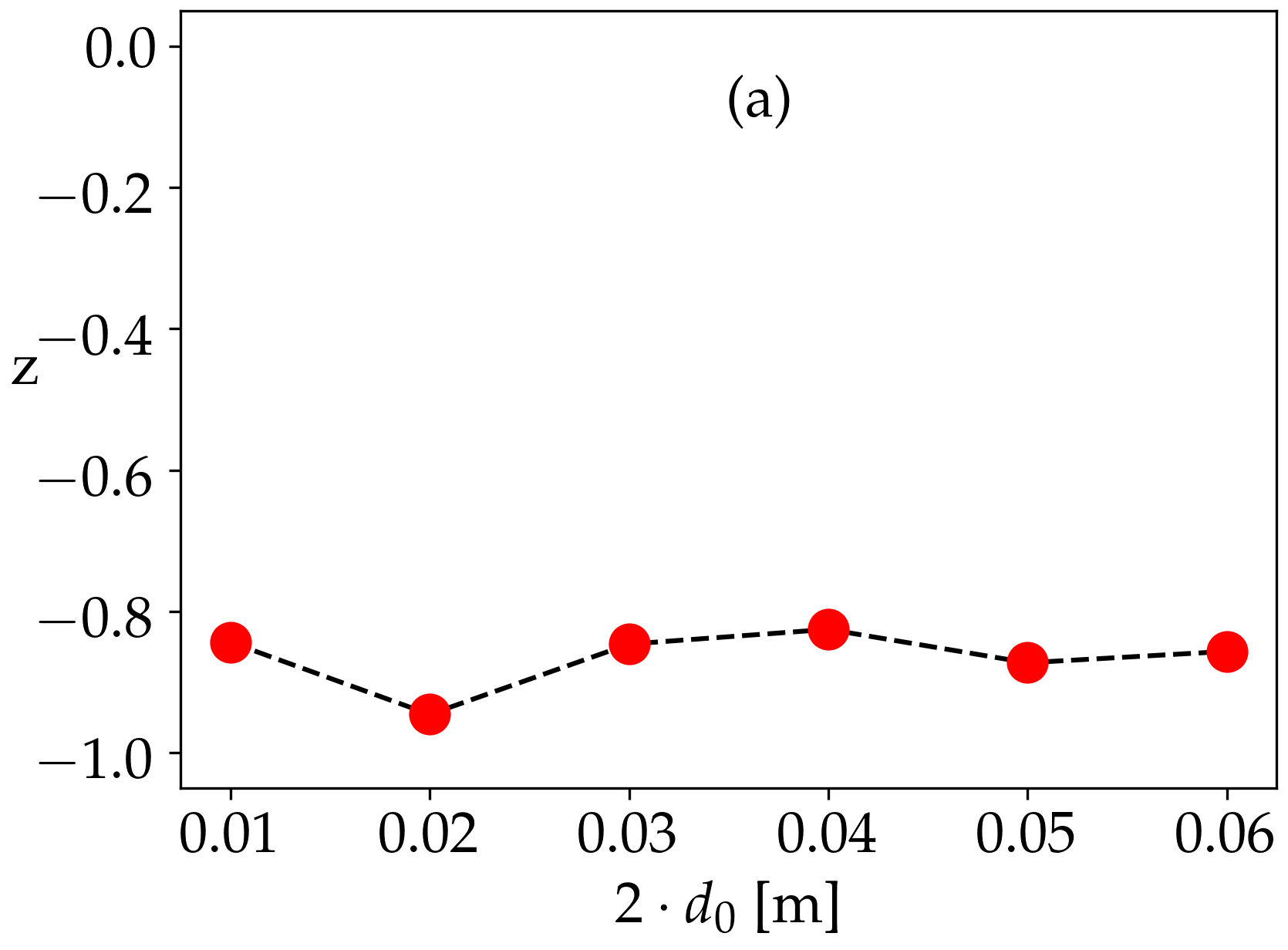}
		\label{fig:f_dist0_b}
	\end{subfigure}
	\begin{subfigure}[b]{0.4\textwidth}
		\centering
		\includegraphics[width=\textwidth]{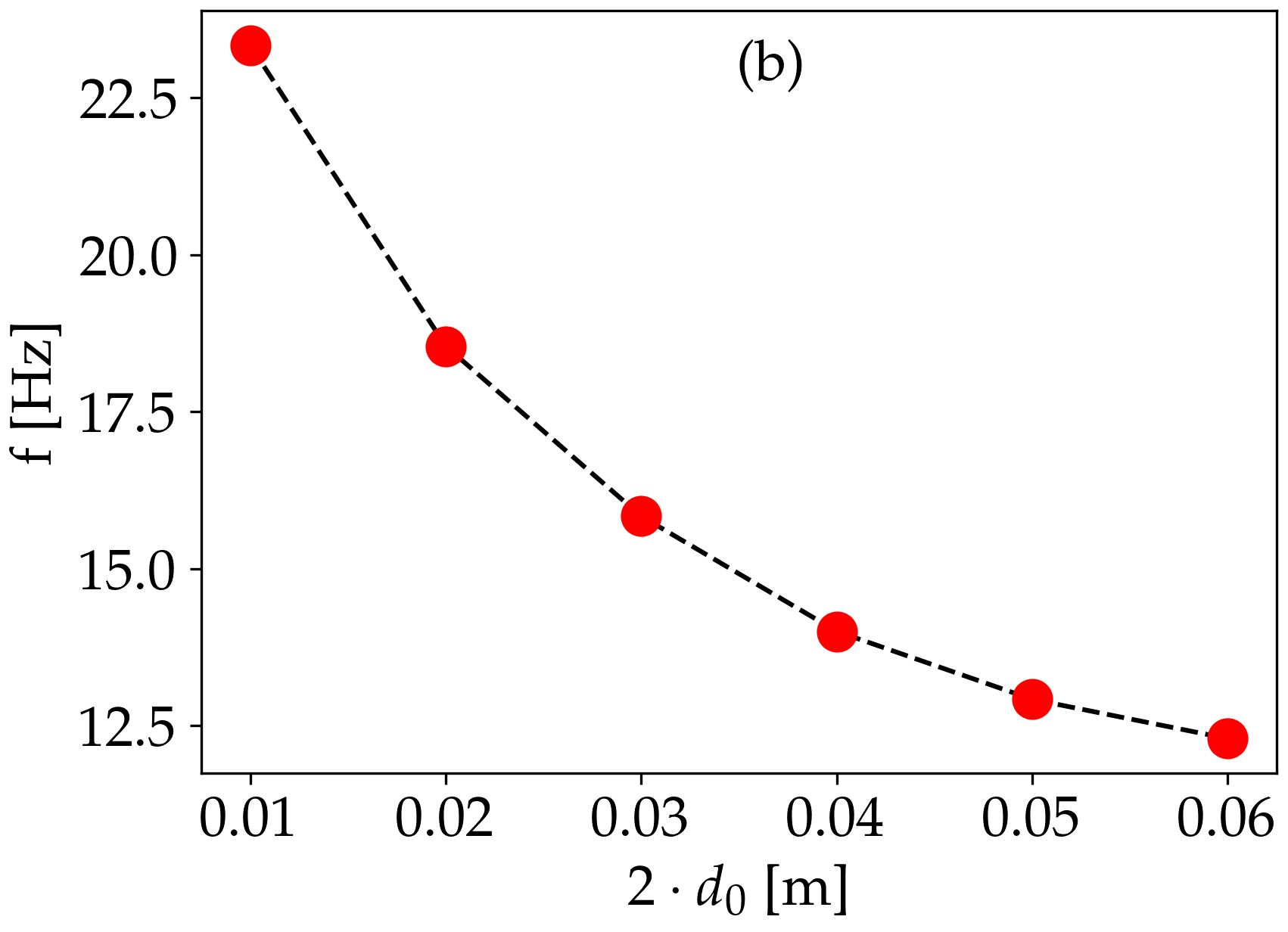}
		\label{fig:z_par_y_a}
	\end{subfigure}
	\caption{Simulation results for the collective behavior of two heated columns. Figure (a)  the synchronization order parameter of two interacting heated fluid columns and Figure (b) shows the collective oscillation frequency, both as a function of the separation distance between the columns. 
The following simulation parameters were considered: 
	$\ \alpha=0.33\cdot 10^{-2}\ K^{-1},\, \rho_0=1.2\ Kg\cdot m^{-2},\, T_0=300\ K,\, D=10^{-4}\ m^2\cdot s^{-1}\cdot K^{-1},\, g_y=-9.81\ m\cdot s^{-2},\, \mu=1.96\cdot10^{-5}\ Kg\cdot s^{-1},\, c_2=0.1\ m\cdot s^{-1},\, c_3=2000\ m^{-1},\, d=0.04\ m,\, c_1=6400\ m^{-1}\cdot s^{-1}$}
	\label{fig:collective small scale}
\end{figure}

For the simulations performed on the smaller length-scale, corresponding to the experimental 
conditions in \cite{Gergely2021} we computed the synchronization order parameter which is meant to characterize the collective oscillation. We used the same $z$ synchronization order parameter as the one used in \cite{Gergely2020} and \cite{Gergely2021}. The computationally derived synchronization parameter is plotted in Figure \ref{fig:collective small scale} a. It's values  in the neighborhood of -1 indicates that we have counter-phase synchronization for the studied distances. 
In Figure \ref{fig:collective small scale}b we also show the oscillation frequency of the two synchronized heated fluid columns as a function of their separation distance. This  frequency decreases as we increase the separation distance between the columns,  similarly with what has been reported in our experiments for the Helium columns \cite{Gergely2021}.

\section{Discussion and Conclusions}

In our previous study \cite{Gergely2021}, we used both experimental and theoretical approaches to investigate whether the hydrodynamic instabilities that occur in rising gas columns is responsible also for the oscillations observed for the diffusion flames \cite{Gergely2020}. It was shown that indeed this is the case: Helium columns ascending in air from a circular nozzle produce similar oscillations with the ones observed in diffusion flames. Also,  the similar collective behavior of these oscillations (counter-phase synchronization) for Helium columns and flickering candle flames suggest that the hydrodynamic processes by their own are enough to explain these phenomenon.

In our previous work \cite{Gergely2021}, for modeling the observed oscillations , a simplified but analytically treatable hydrodynamical approach was used. The model predicted the right trends for the oscillation frequencies as a function of the relevant parameters, but was unsuitable to approach the collective behavior.  In the present follow-up,  we offered an improved modeling by considering a 2D numerical hydrodynamics computer simulation where for computational simplicity instead of ascending Helium columns, heated fluid columns were considered. This approach proved to be successful for reproducing  all the experimentally observed features. For a constant nozzle diameter the numerical hydrodynamics approach lead to an oscillation frequency that increased roughly linearly with the flow yield, in agreement with the experimental results.   For constant flow yield the numerical results suggested a decreasing trend of the oscillation frequency as a function of the nozzle diameter,  confirming the experimental results.  
The exact shape of the simulated trend was slightly different however from the one observed in the experiments. The main reason for this discrepancy is probably the reduction of the real 3D problem to a 2D topology.  Finally, the presented computer simulations were successful also in reproducing the counter-phase synchronization of two heated fluid columns placed nearby each other. The computed  trends for the synchronization order-parameter and the collective frequency were also in agreement
with the experimental results obtained for rising Helium columns in air. 

It worth mentioning here that the computer simulations were performed both for laboratory and for a much larger length-scale than the experiments.  The qualitative agreement between the results (trends and collective behavior)  on these different length-scales suggests that the investigated phenomenon is more general than it was thought to be,  and might have further, yet unexplored connections.

\vspace{2cm}
{\bf Acknowledgment} We acknowledge Dr. M. Ercsey-Ravasz, Dr. B. Moln\'ar, and Prof. dr. G.C. Silaghi, all from the Babes-Bolyai University in helping us with the computational resources for running our simulation codes.

\newpage

\section*{Appendix}

\subsection*{A. Solving the 2D Poison equation in FEniCS}

Here we illsutrate through the 2D Poison equation how to solve a PDE in FEniCS.
The Poisson equation can be given in the following form:
\begin{equation}
-\Delta \varphi = f
\label{eq:poisson}
\end{equation}
If we have a simple rectangular space then the above equation can be easily given in the finite element form 
\begin{equation}
-\frac{\varphi_{i-1,j}-2\varphi_{i,j}+\varphi_{i+1,j}}{h^2}-\frac{\varphi_{i,j-1}-2\varphi_{i,j}+\varphi_{i,j+1}}{h^2} = f_{i,j},
\end{equation}
however, with this simple and intuitive approach, we soon run into problems because even for a circle it is impossible to map the boundary with an acceptably small number of squares. 
The FEniCS program \cite{LangtangenLogg2017} uses a triangular grid instead of a square grid to cover the simulated space, in which case we can always select the grid so that the grid points are on the boundary surfaces.
Discretization alone does not solve the equation, the next question is how to determine the solution at each lattice point.
As a first step, we write the $ \varphi (x_1,x_2)$ function in the following form:
\begin{equation}
 \varphi (x_1,x_2)=\sum_{i=0}^{N}c_{i}\phi_{i}(x_1,x_2)
\end{equation}
In the above equation, $ \phi_{i}(x_1,x_2)$  is a given k-th order polynomial, $c_i$ are the coefficients that determine $\varphi (x_1,x_2)$ and N+1 is the number of the grid points. 
The $c_i$ coefficients are determined by multiplying the Poison equation by N+1 different $v(x_1,x_2)$ so-called test functions and integrating the product over the whole domain to obtain N + 1 linearly independent equations from which the $c_i$ coefficients can be calculated.
All this can be formally given in the following form:
\begin{equation}
    \int_{\Omega} - v \, \Delta \varphi \ d\Omega= \int_{\Omega} f v \ d\Omega
    \label{eq:1}
\end{equation}
We denoted by $d\Omega=dx_1 dx_2$.The above form of PDE is called the weak formulation of the equation and this is what is calculated by the FEniCS program.
The second-order coordinate derivative in the above equation means that the polynomials used need to be twice differentiable. Because the use of polynomials with large degrees requires more memory and computation, we always strive to keep the degree of polynomials to a minimum.
In the above equation, the reduction of the order of derivatives can be done by Gauss-Green integration as follows:
\begin{equation}
    \int_{\Omega}-v\, \Delta \varphi \ d\Omega=  \int_{\Omega} \nabla v \nabla \varphi\ d \Omega- \int_{\partial \Omega} \frac{\partial \varphi}{\partial {\bf n}} v\ d {  s}
\end{equation}
Since we use Dirichlet boundary conditions for the Poisson problem, the value of $v$ at the boundary is 0, so the equation (\ref{eq:1}) can be written in the following form:
\begin{equation}
    \int_{\Omega}\nabla v \nabla \varphi\ d\Omega=\int_{\Omega} f \, v\ d\Omega
\end{equation}
We have seen above how to rewrite the 2D Poisson problem in a form that can be solved with the FEniCS program, now we show the implementation of the solution in Python. 
\begin{lstlisting}[basicstyle=\tiny,language=Python]
from fenics import *	#
import numpy as np #Numpy is required for error calculation
import matplotlib.pyplot as plt #We plot the result with the matplotlib
Nx=10#The number of grid points in the x directions
Ny=10#The number of grid points in the y directions
mesh=UnitSquareMesh(Nx,Ny)
V=FunctionSpace(mesh,'P',1)#space containing first degree polynomials
fi_D=Expression('1+x[0]*x[0]+2*x[1]*x[1]',degree=2)
                    #boundary conditions equation (on boundarys fi(x,y)=x^2+2y^2+1)
def boundery(x,on_boundary):
	return on_boundary
bc=DirichletBC(V,fi_D,boundery)#boundary conditions
fi=TrialFunction(V)
v=TestFunction(V)
f=Constant(-6)
a=dot(grad(fi),grad(v))*dx#right side of equation
L=f*v*dx#left side of equation
fi=Function(V)
solve(a==L,u,bc)#solve the equation

c = plot(interpolate(fi, V), mode='color')
plt.colorbar(c)
plot(fi)
plt.savefig('result1.png')
plt.show()
vertex_v_ud=fi_D.compute_vertex_values(mesh)
vertex_v_u=fi.compute_vertex_values(mesh)
err_max=np.max(np.abs(vertex_v_ud-vertex_v_u))
print("maximum error: ",err_max)
\end{lstlisting}

The above program solves equation (\ref{eq:poisson}) on the unit square of $\{(0,0),(1,1)\}$. The largest difference between the theoretically expected and the numerically obtained value was of the order of the precision of
the numerical representation of the numbers, giving us confidence for the use of the numerical solution. 

\subsection*{B. Test for the 2D hydrodynamics simulations. Karman vortices} 

The first phenomenon we aimed to reproduce using our fluid dynamics simulation is the formation of Karman vortices in the flow of fluids around an obstacle. With this test we aimed to check visually whether the Navier-Stokes equation has been correctly implanted, since the temperature of the fluid at all points is considered fixed: $T_0$. Therefore for this test the density in the simulated volume is constant. 

\begin{figure}
	\centering 
	\includegraphics[width=0.75\textwidth]{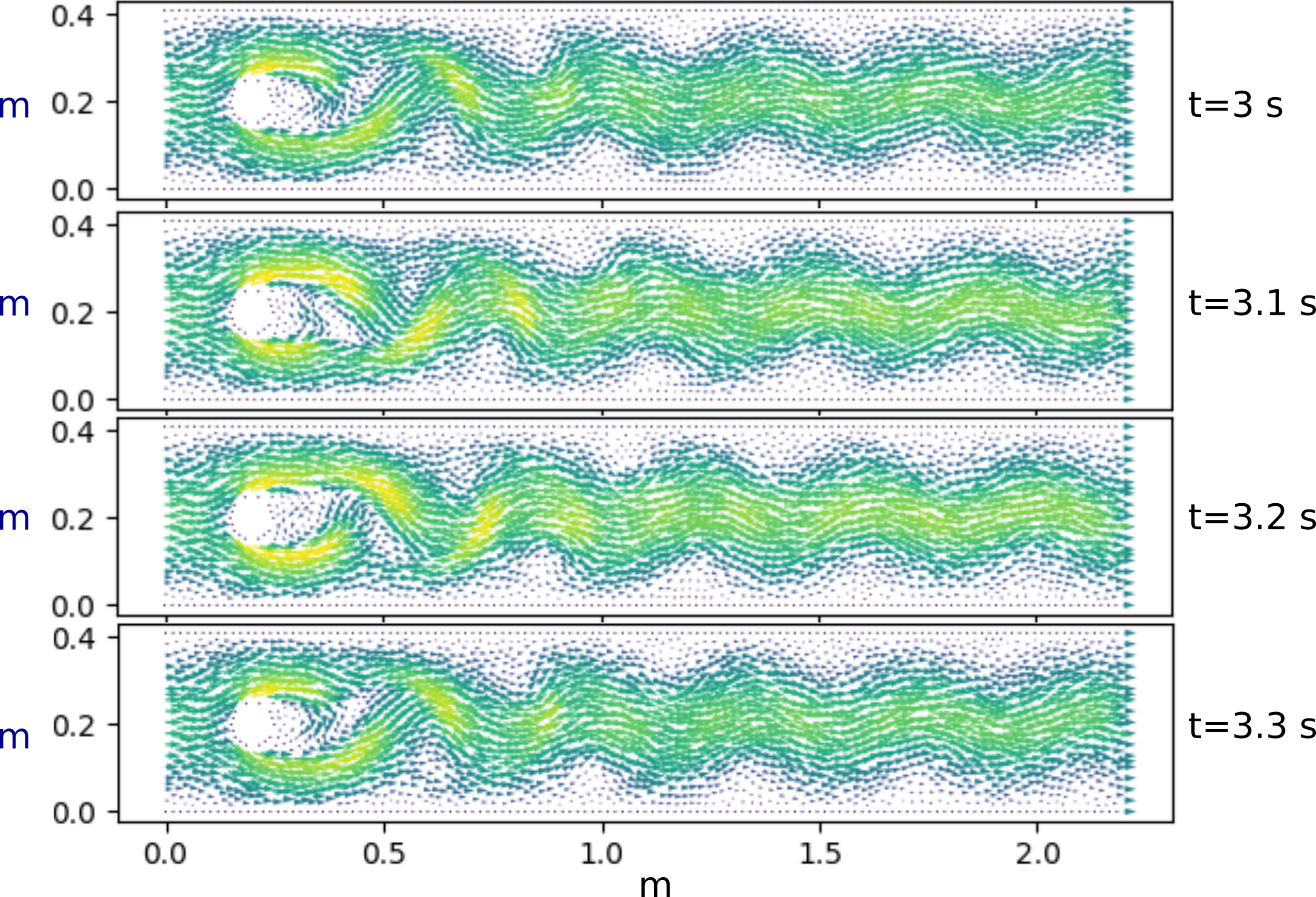}
	\caption{Velocity fields obtained from the simulation of the Karman vortices at four consecutive time moments.}
	\label{fig:Karman}
\end{figure}

In these simulations, the length of the simulated volume was considered as $2.2$m, the width of the simulated volume is $0.41$m  and in the middle of the simulated coordinate space  a circular obstacle with a radius of $0.05$m is placed. The coordinates of the centre of this obstacle was taken at 
(0.2m, 0.2m). 
The density of the fluid was  taken as unity (1kg/m$^2$), the viscosity is $0.001$kg/s, and no gravitational field is considered. For input (the left region of the space in Figure \ref{fig:Karman}) we considered that the velocity in the $y$ direction is $0$, the velocity in the $x$ direction has a parabolic profile with a maximum value of $1.5$m/s. On the horizontal walls and on the boundary of the obstacle we consider no-slip conditions, thus the velocity is fixed to $0$.  For the output (right-sight region) we have also imposed for the $y$ direction velocity to be zero, and the pressure at the output is fixed also to $0$.  Elsewhere there are all free boundary conditions for the pressure. The temperature is fixed at $T_0 = 300$K at input and on the horizontal walls.
 
The velocity vector spaces obtained from the simulations are shown in Figure \ref{fig:Karman} for $4$ time moments as it is indicated on the left side of the images. 
One can observe that the simulation reproduces successfully the expected Karman vortices.

\subsection*{C. Test for the 2D hydrodynamics simulations. Heat induced mushroom cloud}

In this second test we aimed to implement the density and temperature evolution of a heated gas sphere in  gravitational field.  It is expected that the shape of the heated gas will follow the known dynamics of a mushroom cloud in a nuclear explosion.

In the performed simulations, the density at $T_0$ was chosen as unity ($1$kg/m$^2$), the ambient temperature $T_0$ is $300$K, the $\alpha$ parameter  in equation (\ref{temp_eq}) is $\alpha=0.001\ K^{-1}$, the thermal diffusion constant $D$ is $0.3$m$^2$/s,  the initially heated sphere temperature is $600$K, the fluid viscosity is taken as $\mu=0.05$kg/s, the gravitational acceleration is $g_y=-9.81$m/s{$^2$} and the size of the simulated volume is $30$m$^2$ in both the $x$ and $y$ directions.

At the bottom and walls of the simulated space, the velocity is fixed to $0$, at the upper boundary, we fix the $x$ component of the velocity to $v_x=0$. For the pressure, the value of $g \cdot \rho_0 \cdot l$ is fixed for the upper boundary, and free boundary conditions are applied to all the other boundaries. For temperature, free boundary condition is applied at the upper boundary if the $y$ direction component of the velocity is positive otherwise the temperature is fixed to $T_0=300$K units on the upper boundary and on the side-walls.  The temperature is fixed to a higher value of  $450$K units at the bottom-wall of the simulated area.  This is necessary in order to make the resulting flow visible in the temperature space. Initially, the velocity in the whole simulated volume is $0$ and the volume contains a sphere (disk)  with a radius of $5$m in which the temperature is $600$K. The centre of the sphere is at the coordinates 
(0m, 7m), the temperature around the sphere is fixed to $300$K, and the temperature between the centre and the surface of the disk is given by an interpolation with a  sigmoid function.
 
\vspace{0.5cm}

\begin{figure}
	\centering 
	\includegraphics[width=0.75\textwidth]{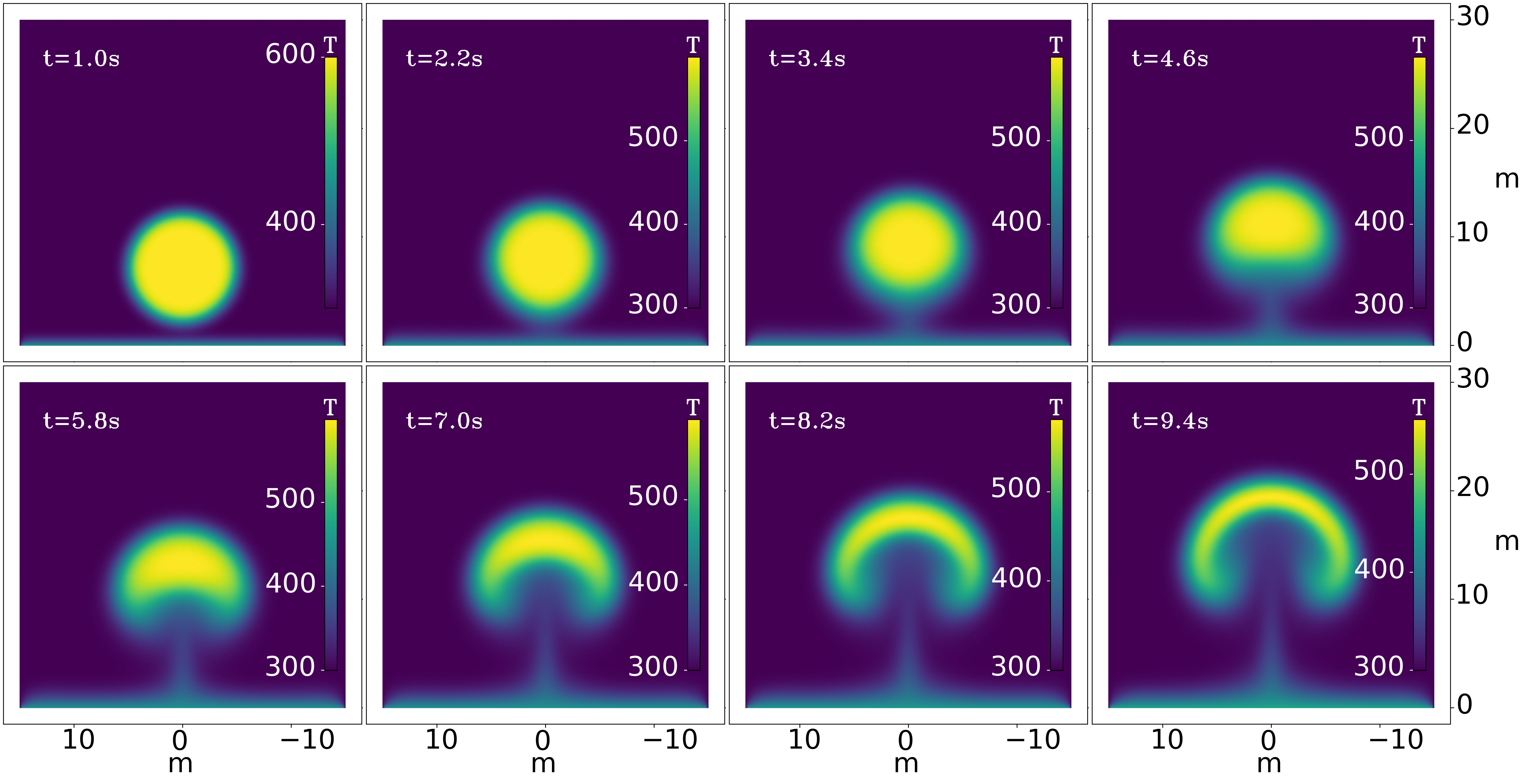}
	\caption{Snapshots of the evolution of a heated gas sphere in time. The parameters and details of the simulation can be found in the text.}
	\label{fig:mushroomC}
\end{figure}

The time-evolution of the temperature map derived from the simulation is shown in Figure \ref{fig:mushroomC}. The effect of thermal diffusion can be observed in the first two frames. As a result of this diffusion the initially sharp boundary line between the high and low temperature regions gets blurred. Subsequent frames show the displacement due to convective flow and as a result of this the characteristic mushroom cloud shape is formed.

\newpage

\bibliography{hydro-numerics}

\end{document}